\documentclass[%
 reprint,
superscriptaddress,
 amsmath,amssymb,
 aps,
 longbibliography,
 floatfix,
]{revtex4-2}

\usepackage{graphicx}% Include figure files
\usepackage{dcolumn}% Align table columns on decimal point
\usepackage{bm}% bold math
\usepackage{braket}
\usepackage{physics}
\usepackage{amsmath}
\usepackage{xcolor}
\usepackage{hyperref}% add hypertext capabilities
\usepackage{qcircuit}

\renewcommand{\vec}[1]{\boldsymbol{#1}}

\newcommand{\dqc}{Duke Quantum Center, Duke University, Durham, NC 27701, USA}

\begin{document}

\preprint{APS/123-QED}

\title{Designing Filter Functions of Frequency-Modulated Pulses for High-Fidelity Two-Qubit Gates in Ion Chains}

\author{Mingyu Kang}
\email{mingyu.kang@duke.edu}
\affiliation{\dqc}
\affiliation{Department of Physics, Duke University, Durham, NC 27708, USA}
\author{Ye Wang}
\email{wang.ye.phy@gmail.com}
\affiliation{\dqc}
\affiliation{Department of Electrical and Computer Engineering, Duke University, Durham, NC 27708, USA}
\author{Chao Fang}
\affiliation{\dqc}
\affiliation{Department of Electrical and Computer Engineering, Duke University, Durham, NC 27708, USA}
\author{Bichen Zhang}
\affiliation{\dqc}
\affiliation{Department of Electrical and Computer Engineering, Duke University, Durham, NC 27708, USA}
\author{Omid Khosravani}
\affiliation{\dqc}
\affiliation{Department of Electrical and Computer Engineering, Duke University, Durham, NC 27708, USA}
\author{Jungsang Kim}
\affiliation{\dqc}
\affiliation{Department of Physics, Duke University, Durham, NC 27708, USA}
\affiliation{Department of Electrical and Computer Engineering, Duke University, Durham, NC 27708, USA}
\affiliation{IonQ, Inc., College Park, MD 20740, USA}
\author{Kenneth R. Brown}%
\email{ken.brown@duke.edu} 
\affiliation{\dqc}
\affiliation{Department of Physics, Duke University, Durham, NC 27708, USA}
\affiliation{Department of Electrical and Computer Engineering, Duke University, Durham, NC 27708, USA}
\affiliation{Department of Chemistry, Duke University, Durham, NC 27708, USA}

\date{\today}

\begin{abstract}
High-fidelity two-qubit gates in quantum computers are often hampered by fluctuating experimental parameters. The effects of time-varying parameter fluctuations lead to coherent noise on the qubits, which can be suppressed by designing control signals with appropriate filter functions. Here, we develop filter functions for M\o{}lmer-S\o{}rensen gates of trapped-ion quantum computers that accurately predict the change in gate error due to small parameter fluctuations at any frequency. We then design the filter functions of frequency-modulated laser pulses, and compare this method with pulses that are robust to static offsets of the motional-mode frequencies. Experimentally, we measure the noise spectrum of the motional modes and use it for designing the filter functions, which improves the gate fidelity from 99.23(7)\% to 99.55(7)\% in a five-ion chain.
\end{abstract}

\maketitle

\section{Introduction} \label{sec:sec1}

Generating high-fidelity entangling gates in multiqubit systems is a key challenge for scalable quantum computing. Trapped-ion systems with exactly two ions have achieved two-qubit gate fidelities higher than 99.9\%, using lasers \cite{Ballance16, Gaebler16, Clark21} and magnetic field gradients \cite{Srinivas21}. Larger systems, despite remarkable experimental efforts, are more susceptible to various noise and parameter drifts, which makes them more challenging to achieve high-fidelity two-qubit gates. Two-qubit gate fidelities of approximately 99\% for a 15-ion chain \cite{Egan21} and 97.5\% for 16-ion and 25-ion chains \cite{Postler22, Cetina22} have been reported.  

Trapped-ion qubits are entangled by a state-dependent force that briefly excites the normal modes of the ions' collective motion. At the end of the gate, all motional modes should be completely disentangled from the qubits, while the qubit states are entangled with each other by the correct amount \cite{Molmer99, Sorensen99}. Such precise control needs to be performed in the presence of experimental noise.

To achieve this task of robust high-fidelity gates, various pulse-design methods have been proposed. One approach is to use multichromatic beams with tunable amplitudes \cite{Haddadfarshi16, Webb18, Shapira18, Zarantonello19, Shapira20, Blumel21P}, and another approach is to use amplitude \cite{Zhu06, Roos08, Kim09, Choi14, Figgatt19, Blumel21E}, phase \cite{Hayes12, Lu19, Green15, Milne20, Bentley20}, and/or frequency \cite{Wang20, Leung18, Landsman19, Kang21} modulation over many time segments. These methods find a pulse that guarantees high fidelity in the presence of a small offset of a parameter, such as the motional-mode frequency~\cite{Valahu22}.  

While it is useful to achieve robustness to \textit{static} offsets, experimental parameters often fluctuate over time. Recently there has been exciting works in experimentally measuring the noise spectrum of the motional modes \cite{Milne21, Keller21} as well as control signal \cite{Frey17, Nakav22} and ambient dephasing \cite{Wang17, Frey20} in a trapped-ion system. This motivates designing pulses that are robust to \textit{time-varying} parameter fluctuations of a known spectrum. 

The filter-function (FF) formalism describes the performance of a control protocol in the presence of time-varying noise \cite{Kofman01, Biercuk11, Green13, PazSilva14}. In particular, designing the FF has been experimentally shown to be useful for suppressing errors of single-qubit gates in trapped-ion systems \cite{Timoney08, Soare14, Ball21}. The FF for two-qubit gates has been introduced in Refs.~\cite{Green15, Milne20, Ball21}, but has limited capability in predicting the response of the gate error to noise of frequency lower than the inverse of gate time \cite{Milne20}. 

We propose a method of actively designing the FFs of frequency-modulated (FM) pulses for two-qubit gates with trapped ions, such that the effects of noise of a given spectrum, including its low-frequency component, are suppressed. The rest of the paper is organized as follows. In Sec.~\ref{sec:sec2} we briefly review the theory of M\o{}lmer-S\o{}rensen (MS) gates and their FFs. In particular, we introduce the FF for the rotation angle with respect to the entangling spin axis, which is crucial for describing the gate error with low-frequency noise. In Sec.~\ref{sec:sec3}, we improve on the previous FM pulse-design scheme \cite{Leung18} by designing the FFs, which lowers the gate error in the presence of time-varying fluctuations as well as static offsets in the motional-mode frequencies. In Sec.~\ref{sec:sec4}, we experimentally demonstrate measuring the noise spectrum and applying the results to designing the FFs, which improves the two-qubit gate fidelity from 99.23(7)\% to 99.55(7)\% in a five-ion chain for a fixed pulse length. Finally, we discuss future directions and summarize our results in Sec.~\ref{sec:sec5}.

%%%%%%%%%%%%%%%%%%%%%%%%%%%%%%%%%%%%%%%%%%%%%%%%%%%%%%%%%%%%%%%%%%%%%%%%5

\section{MS Gate and Filter Functions} \label{sec:sec2}

\subsection{MS-gate errors} \label{sec:subsec2A}

The MS gate using FM pulse applies a state-dependent force with lasers at a drive frequency modulated near the sideband frequencies. As the pulse is applied to ions $j_1$ and $j_2$, the unitary evolution of the system of the ions and the motional modes, after applying the rotating-wave approximation, is given by 
\begin{align}
    \hat{U} = \exp &\Big\{\sum_{j=j_1,j_2}\sum_k 
    \Big([\alpha_{kj} \hat{a}_k^\dagger - \alpha^*_{kj} \hat{a}_k] \:\hat{\sigma}^x_j\Big) + i\Theta \:\hat{\sigma}^x_{j_1} \hat{\sigma}^x_{j_2} \Big\}, \nonumber
\end{align}
where $\hat{a}^\dagger_k$ is the creation operator of mode $k$ and $\hat{\sigma}_j^x$ is the bit-flip operator of ion $j$. Also, $\alpha_{kj}$ is the displacement of motional mode $k$ with respect to ion $j$ and $\Theta$ is the rotation angle of the spins with respect to the $XX$ axis \cite{Wu18}, which are given by
\begin{gather}
    \alpha_{kj} = \frac{\Omega \eta_{kj}}{2} \int^\tau_0 \: e^{-i\theta_k(t)}dt, \label{eq:alpha}\\
    \Theta = - \Omega^2 \sum_k \frac{\eta_{kj_1} \eta_{kj_2}}{2} \int^\tau_0 dt_1
    \int^{t_1}_0 dt_2 \sin [\theta_k(t_1) - \theta_k(t_2) ], \label{eq:Theta}
\end{gather}
where
\begin{equation}
    \theta_k(t) = \int^t_0 [\mu(t') - \omega_k] dt' . \label{eq:theta}
\end{equation}
Here, $\tau$ is the pulse length, $\Omega$ is the carrier Rabi frequency, $\eta_{kj}$ is the Lamb-Dicke parameter of ion $j$ with respect to mode $k$, and $\omega_k$ is the frequency of mode $k$. Also, $\mu(t)$ is the drive frequency, which we call the FM pulse.  

An ideal MS gate satisfies $\alpha_{kj_1} = \alpha_{kj_2} = 0$ $\forall k$ and \mbox{$\Theta = \pi/4$}, where the first condition is necessary to completely disentangle the qubits from the motional modes at the gate's conclusion. The two-qubit-gate error $\mathcal{E}$, defined as the normalized Hilbert-Schmidt inner product of the ideal and actual unitary operators, can be expressed up to leading order as $\mathcal{E} = \mathcal{E}_\alpha + \mathcal{E}_\Theta$, where $\mathcal{E}_\alpha$ and $\mathcal{E}_\Theta$ are, respectively, the displacement and angle error \cite{Bentley20}, given by
\begin{align}
    \mathcal{E}_\alpha &= \sum_k \left(|\alpha_{kj_1}|^2 + |\alpha_{kj_2}|^2 \right), \label{eq:disp_err}\\
    \mathcal{E}_\Theta &= \left(\Theta - \frac{\pi}{4}\right)^2. \label{eq:angle_err}
\end{align}
Note that here we assume zero temperature. At higher temperature, such that the initial average phonon occupation of mode $k$ is $\bar{n}_k$, the contribution of mode $k$ to $\mathcal{E}_\alpha$ has an additional proportionality of $(\bar{n}_k+\frac{1}{2})$~\cite{Bentley20, Kang21}. Using sideband cooling, $\bar{n}_k \lesssim 0.1$ is achievable~\cite{Monroe95}, and motional heating rate of the near-resonantly excited modes can be maintained below approximately $2 \times 10^{-3}$ quanta per gate time~\cite{Wang20}, which make the zero-temperature approximation valid.

\subsection{Robustness to static mode-frequency offsets} \label{sec:subsec2B}

In this paper we focus on the static offsets and time-varying fluctuations of the mode frequencies $\omega_k$, which occur from various classical sources of noise, such as the fluctuation of the rf driving signal for the trap. This is motivated by the fact that motional dephasing is one of the leading sources of errors for MS gates in our system~\cite{Wang20}. The effects of fluctuations in other parameters, such as the laser phase and intensity, are explored in Appendix~\ref{app:Phase} and \ref{app:Laser}.

First, we consider the effects of \textit{static} mode-frequency offsets on the displacement error. When \mbox{$\omega_k \rightarrow \omega_k + \delta_k$}, where $\delta_k$ is the unwanted mode-frequency offset, Eq.~\ref{eq:disp_err} becomes
\begin{equation} \label{eq:staticdisperr}
    \mathcal{E}_\alpha = \sum_{j=j_1,j_2} \sum_k \left| \sum_{m=0}^\infty \frac{\delta_k^m}{m!} \frac{\partial^m \alpha_{kj}}{\partial \omega_k^m} \right|^2,
\end{equation}
for sufficiently small $\delta_k$'s such that the sum converges. Therefore, an analytic approach of minimizing $\mathcal{E}_\alpha$ for any $\delta_k$'s, up to order $M$, is to minimize $|\partial^m \alpha_{kj} / \partial \omega_k^m|$ to zero for all $k$ and $m=0, 1, .., M$ \cite{Blumel21P}. 

Note that the first-order derivative can be expressed as~\cite{Leung18}
\begin{equation}
    \frac{\partial \alpha_{kj}}{\partial \omega_k} = i \tau \left(\alpha_{kj} - \bar{\alpha}_{kj}\right), \nonumber
\end{equation}
where
\begin{equation}\label{eq:avgdisp}
    \bar{\alpha}_{kj} = \frac{\Omega \eta_{kj}}{2\tau} 
    \int^\tau_0 \int_0^t e^{-i \theta_k(t')}dt' dt
\end{equation}
is the time-averaged displacement. Thus, a first-order robustness \mbox{($M = 1$)} can be achieved by minimizing $|\alpha_{kj}|$ and $|\bar{\alpha}_{kj}|$ to zero for all $k$. This is convenient as using a time-symmetric pulse [$\mu(\tau-t) = \mu(t)$] guarantees that minimizing $|\bar{\alpha}_{kj}|$ also minimizes $|\alpha_{kj}|$~\cite{Leung18}. 

Next, we consider the angle error. For simplicity we assume $\delta_k = r_k \delta$ $\forall k$, i.e. offsets of different modes differ only up to proportionality constants. This is a valid assumption when $\delta_k(t)$ comes from noise in the trap's rf voltage. Then, we may consider the $m$th-order derivative of $\Theta$ over $\delta$. Specifically, for $m=1$,
\begin{align}
\frac{\partial \Theta}{\partial \delta} &= \Omega^2 \sum_k \frac{r_k}{2} \eta_{kj_1} \eta_{kj_2} \nonumber\\
&\quad \times \int^\tau_0 dt_1 \int^{t_1}_0 dt_2  (t_1 - t_2) \cos [\theta_k(t_1) - \theta_k(t_2) ],
\label{eq:anglederivative}
\end{align}
where the derivative is evaluated at $\delta = 0$. For the robustness of the rotation angle, $|\partial \Theta / \partial \delta|$ needs to be minimized, on top of satisfying the usual condition $\Theta = \pi/4$.

\subsection{Robustness to time-varying mode-frequency fluctuations} \label{sec:subsec2C}

The effects of \textit{time-varying} mode-frequency fluctuations $\delta_k(t)$ on the gate errors can be described using the FF formalism. For simplicity we assume $\delta_k(t) = r_k \delta(t)$. The power spectral density (PSD) of $\delta(t)$ is defined as
\begin{equation} \label{eq:Sdef}
    S_{\delta}(f) = \int_{-\infty}^\infty dt \: 
    \langle \delta(t'+t)\delta(t') \rangle \: e^{-2 \pi i f t},
\end{equation}
where $\langle \cdot \rangle$ denotes the average over all $t'$. 

\begin{figure}[ht]
\includegraphics[width=8.6cm]{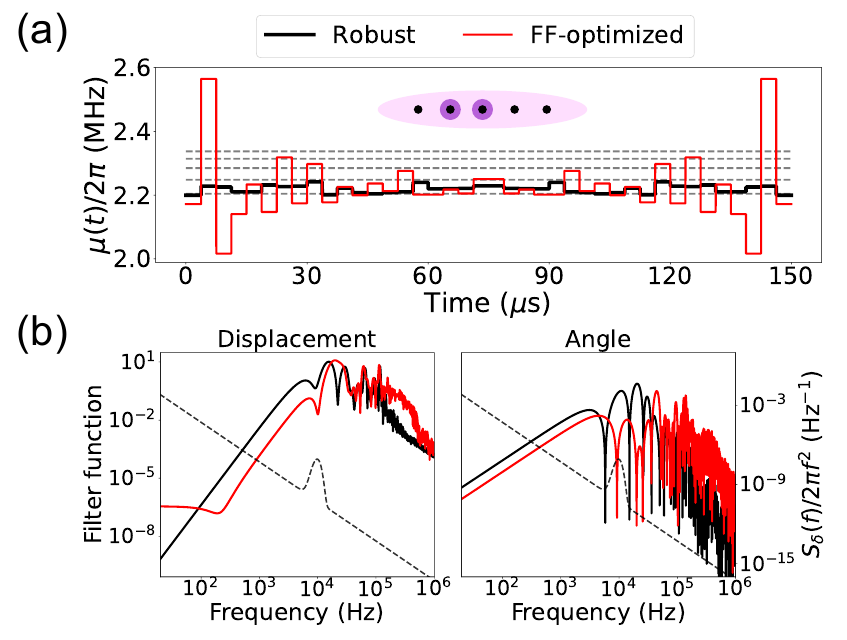}
\caption{(a) Pulses obtained by robust FM (black) and FF optimization (red), which require carrier Rabi frequency $\Omega/2\pi = 85.8$ and $109.8$ kHz, respectively. 150-$\mu$s pulses are applied on the second and third ions of a five-ion chain with the sideband frequencies shown as dashed lines. (b) Filter functions $F_\alpha(f)$ (left) and $F_\Theta(f)$ (right) of the pulses. Dashed lines show the noise spectrum $S_\delta(f)/f^2$ with the characteristic frequency $f_c = 10$ kHz that is used in the FF optimization.}
\label{fig:pulses}
\end{figure}

Now we introduce the FFs for MS gates. Consider the case where a pulse satisfies the ideal conditions \mbox{$\alpha_{kj_1} = \alpha_{kj_2} = 0$} $\forall k$ and \mbox{$\Theta = \pi/4$}, but gate error occurs due to fluctuations \mbox{$\omega_k \rightarrow \omega_k + r_k \delta(t)$}. Then, for small fluctuations such that \mbox{$|\int_0^t \delta_k(t')dt'| \ll 1 \: (0 \leq t \leq \tau)$}, up to leading order, the errors are given by
\begin{equation}
    \mathcal{E}_\nu = \int_{-\infty}^\infty df \:\frac{S_{\delta}(f)}{f^2} F_{\nu} (f) \label{eq:SF},
\end{equation}
where $\nu = \alpha$ or $\Theta$, and 
\begin{align}
    F_\alpha(f) &= \Omega^2 \sum_k (\eta_{kj_1}^2 + \eta_{kj_2}^2)
                  \Big| \frac{r_k}{2} \int_0^\tau dt \:e^{i(2\pi f t - \theta_k(t))} \Big|^2, \label{eq:disp_ff}\\
    F_\Theta(f) &= \Omega^4 \Big|\int^\tau_0 dt_1 \int^{t_1}_0 dt_2 \: 
    (e^{2\pi i f t_1} - e^{2 \pi i f t_2}) \nonumber\\
    &\quad\quad\quad\quad\times \sum_k \frac{r_k}{2}\: \eta_{kj_1} \eta_{kj_2} \cos [\theta_k(t_1) - \theta_k(t_2)] \: \Big|^2. \label{eq:angle_ff}
\end{align}
Here, $S_\delta (f)/f^2$ is the noise spectrum of motional dephasing, and $F_\alpha(f)$ and $F_\Theta(f)$ are the FFs for the displacement and angle error, respectively. Therefore, given the noise spectrum, the gate error can be minimized by designing the FFs. Figure~\ref{fig:pulses} shows an example. The method used in designing the FFs here will be described in Sec.~\ref{sec:subsec3A}.

%%%%%%%%%%%%%%%FIGURE 1%%%%%%%%%%%%%%
%%%%%%%%%%%%%%%%%%%%%%%%%%%%%%%%%%%%

Note that while the previous literature considers only $F_\alpha(f)$ \cite{Green15, Milne20, Ball21}, $F_\Theta(f)$ is larger than $F_\alpha(f)$ at frequencies $f \ll 1/\tau$, so is crucial for minimizing the gate error in the presence of low-frequency noise. The derivation of $F_\Theta(f)$ is given in Appendix~\ref{app:Derivations}. 

The usage of FFs is not limited to handling mode-frequency fluctuations. Appendix~\ref{app:Phase} shows that for an experimental setup where fluctuations of the \textit{motion phase} of lasers becomes relevant~\cite{Lee05}, $F_\alpha(f)$ and $F_\Theta(f)$ can be used for laser-phase noise as well. Appendix~\ref{app:Laser} introduces a different set of FFs that describes the effects of laser-intensity noise, or fluctuations of $\Omega$.

\subsection{FFs and static mode-frequency offsets} \label{sec:subsec2D}

While the FFs above are designed to handle time-varying parameter fluctuations, it can be easily shown that suppressing the FFs at low frequencies leads to reducing the gate error due to static mode-frequency offsets as well. 

First, we consider the displacement FF. Expanding $e^{2\pi i f t}$ in Eq.~\ref{eq:disp_ff} into a Taylor series gives
\begin{equation} \label{eq:staticFF}
F_\alpha(f) = \sum_{j=j_1,j_2} \sum_k \left| r_k \sum_{m=0}^\infty \frac{(2\pi f)^m}{m!} \frac{\partial^m \alpha_{kj}}{\partial \omega_k^m} \right|^2,
\end{equation}
which closely resembles the displacement error with static offsets in Eq.~\ref{eq:staticdisperr}. In particular, when $r_k = 1$, $F_\alpha(f)$ is equal to the displacement error when $\delta_k = 2\pi f$ $\forall k$. Thus, we expect that suppressing $F_\alpha(f)$ at low frequencies leads to reducing the displacement error due to small static offsets. 

Note that Eq.~\ref{eq:staticdisperr} is valid only for small $\delta_k$'s such that the sum converges. Therefore, the predictions of FFs become inaccurate when the noise corresponds to large static offsets during the gate time. This agrees with previous observations that the first-order approximation of the FF formalism is less accurate when noise is stronger at frequencies lower than the inverse of the gate time~\cite{Kabytayev14, Milne20}.

Next, we consider the angle FF. Comparing Eqs. \ref{eq:anglederivative} and \ref{eq:angle_ff}, we immediately obtain
\begin{equation}
\lim_{f \rightarrow 0} \frac{F_\Theta(f)}{f^2} = (2\pi)^2 \left| \frac{\partial \Theta}{\partial \delta} \right|^2,
\end{equation}
which shows that suppressing $F_\Theta(f)$ at low frequencies reduces the angle error due to small static offsets. Unlike the case of the displacement, $F_\Theta(f)$ at low frequency is explicitly related to only the first-order derivative of $\Theta$, and not to its higher-order derivatives.

%%%%%%%%%%%%%%%%%%%%%%%%%%%%%%%%%%%%%%%%%%%%%%%%%%%%%%%%%%%%%%%

\section{Comparison of Optimization Methods} \label{sec:sec3}

%%%%%%%%%%%%%%%FIGURE 2%%%%%%%%%%%%%%
\begin{figure*}[ht!]
\includegraphics[width=18.5cm]{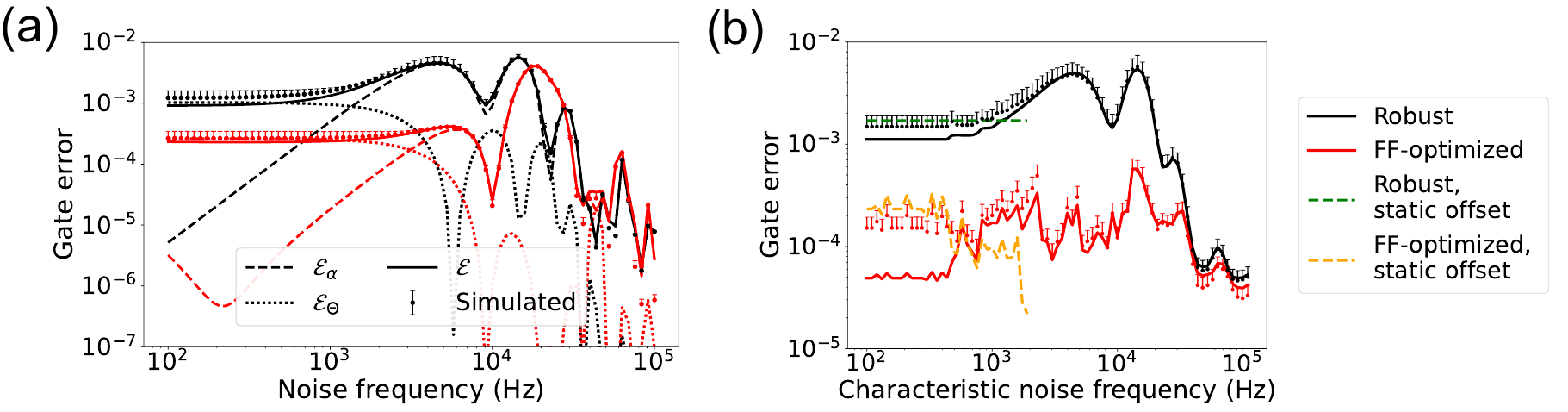}
\caption{Gate errors of the 150-$\mu$s pulses obtained by robust FM (black) and FF optimization (red), predicted (lines) by Eqs.~\ref{eq:SF}-\ref{eq:angle_ff} and simulated (dots) by state-vector evolution. (a) The pulses in Fig.~\ref{fig:pulses} are compared under monotone fluctuations of various frequencies injected to mode frequency $\omega_k$. The amplitude of fluctuation is fixed to \mbox{$2\sqrt{2}\pi (\omega_k / \omega_{\rm CM}) \times 500$ Hz}. Each error bar represents the upper standard deviation of the simulated gate errors over 1000 initial phases of the fluctuation. (b) For each $S_\delta(f)$, defined with the characteristic noise frequency $f_c$ by Eqs.~\ref{eq:Sdelta}-\ref{eq:Sdelta2}, we find the FF-optimized pulse, which requires carrier Rabi frequency $\Omega/2\pi$ between 67 and 150 kHz, and compare the gate error with the robust-FM pulse shown in Fig.~\ref{fig:pulses}. For the simulations, fluctuations $\delta(t)$ generated from each $S_\delta(f)$ are injected to the mode frequencies. Each error bar represents the upper standard deviation of the simulated gate errors over 1000 realizations of $\delta(t)$. The dashed lines show the average predicted gate errors when static offsets, drawn from a normal distribution of zero mean and standard deviation \mbox{$2\pi \times (500^2 + 100^2)^{1/2}$ Hz}, are added to the mode frequencies.}
\label{fig:timevarying}
\end{figure*}
%%%%%%%%%%%%%%%%%%%%%%%%%%%%%%%%%%%%

\subsection{Optimization methods} \label{sec:subsec3A}

In this subsection we introduce the FM pulse-optimization methods that achieve robustness to static offsets and time-varying fluctuations of the mode frequencies, where the latter is done by our ``FF-optimization'' method. 

The previous FM pulse-design method, which we call ``robust FM'' \cite{Leung18}, finds a pulse $\mu(t)$ that removes the time-averaged displacement. The cost function is given by
\begin{equation}
    C_1 = \sum_k \left( |\bar{\alpha}_{kj_1}|^2 + |\bar{\alpha}_{kj_2}|^2 \right).
\end{equation}
As explained in Sec.~\ref{sec:subsec2B}, by constraining $\mu(t)$ to time-symmetric pulses, minimizing $C_1$ removes the displacement of each mode up to first order ($M=1$) in the static mode-frequency offset. 

We may also consider the ``second-order robust-FM'' method, which finds $\mu(t)$ that minimizes the displacement up to second order ($M=2$), using the cost function
\begin{equation}
    C_2 = C_1 + \frac{1}{\tau^2} \sum_k \left(
    \left| \frac{\partial^2 \alpha_{kj_1}}{\partial \omega_k^2} \right|^2 
    + \left| \frac{\partial^2 \alpha_{kj_2}}{\partial \omega_k^2} \right|^2
    \right).
\end{equation}

Finally, we introduce the ``FF-optimization'' method, which designs the FFs $F_\alpha(f)$ and $F_\Theta(f)$ for a given noise PSD $S_{\delta}(f)$. Specifically, we find $\mu(t)$ that minimizes the cost function
\begin{equation}
C_{\rm FF} = C_1 + \int_{-f_{\rm max}}^{f_{\rm max}} df \frac{S_\delta(f)}{f^2} \left[F_\alpha(f) + F_\Theta(f) \right], \label{eq:costfn}
\end{equation}
where $f_{\rm max}$ is the cutoff frequency. In this paper, we choose $f_{\rm max} = 13.3$ MHz  as we consider $S_\delta(f)$ that decays at high frequencies. 

For all methods above, we constrain $\mu(t)$ to piecewise-constant and time-symmetric pulses. The gradient of each term in the cost function over each segment of $\mu(t)$ is analytically evaluated for efficient optimization. Also, $\Omega$ is updated at each iteration of optimization such that \mbox{$\Theta = \pi/4$} in the absence of noise. 

In this paper, unless specified otherwise, ``robust FM'' indicates the first-order method. For the majority of the paper, we compare the (first-order) robust-FM and the FF-optimization methods. The second-order robust-FM method will be briefly used in Sec.~\ref{sec:subsec3C} to discuss robustness to static mode-frequency offsets in detail. 

For the FF optimization performed in this section, in order to study the noise of various frequencies, we consider $S_\delta (f)$ that consists of two types of noise: (i) noise of a Gaussian spectrum centered at the characteristic frequency $f_c$, and (ii) $1/f$ noise. Specifically, $S_\delta (f)$ is given by
\begin{equation}\label{eq:Sdelta}
    S_{\delta}(f) = \big(S_{\delta,1}(f)^{1/2} + S_{\delta,2}(f)^{1/2}\big)^2,
\end{equation}
where
\begin{align}
    S_{\delta,1}(f) &= \frac{\mathcal{N}_1}{\sqrt{2\pi}\sigma}
    \exp\big(-\frac{(f - f_c)^2}{2\sigma^2}\big), \label{eq:Sdelta1}\\
    S_{\delta,2}(f) &= \frac{\mathcal{N}_2}{f}. \label{eq:Sdelta2}
\end{align}
For each $f_c$, we set $\sigma = f_c / 10$, and choose $\mathcal{N}_1$ such that the standard deviation of $\delta(t)$ realized from $S_{\delta,1}(f)$ is \mbox{$2\pi \times 500$ Hz}. Also, $\mathcal{N}_2$ is chosen such that the standard deviation of $\delta(t)$ realized from $S_{\delta,2}(f)$ is \mbox{$2\pi \times 100$ Hz}.

Figure~\ref{fig:pulses} shows the robust-FM and FF-optimized pulses and their FFs $F_\alpha(f)$ and $F_\Theta(f)$. We use 150-$\mu$s pulses to perform a MS gate on the second and third ions of a five-ion chain. For the FF optimization, we use $S_\delta (f)$ in Eqs.~\ref{eq:Sdelta}-\ref{eq:Sdelta2} with $f_c = 10$ kHz, which is shown as dashed lines in Fig.~\ref{fig:pulses}(b). Also, we use $r_k = \omega_k / \omega_{\rm CM}$, where $\omega_{\rm CM}$ is the frequency of the center-of-mass mode (largest $\omega_k$), which is a reasonable assumption for the rf-voltage fluctuations. 

In Fig.~\ref{fig:pulses}(b), $F_\alpha(f)$ of the robust-FM pulse converges to zero as $f \rightarrow 0$. This is because minimizing the displacement $|\alpha_{kj}|$ to zero also minimizes $F_\alpha(0)$ to zero. By relaxing this constraint, such that small displacements in the absence of noise are allowed \cite{Kang21}, the FF-optimization method is able to find a pulse that suppresses both $F_\alpha(f)$ and $F_\Theta(f)$ at frequencies near and lower than $f_c$.

\subsection{Comparison under time-varying fluctuations} \label{sec:subsec3B}

To test whether the FFs can accurately describe the gate error, we inject monotone fluctuation of frequency $f'$ into the mode frequencies, and compare the gate error predicted by Eqs.~\ref{eq:SF}-\ref{eq:angle_ff} and simulated using Qutip \cite{Qutip}, for various values of $f'$. We use the robust-FM and FF-optimized pulses in Fig.~\ref{fig:pulses}. The simulations are performed by solving the state-vector evolution with respect to the Hamiltonian of the MS gate~\cite{Wu18}.

Figure~\ref{fig:timevarying}(a) shows that the FF-optimized pulse has lower gate error than the robust-FM pulse with any noise of frequency lower than \mbox{17 kHz}. This shows that the gate error can be reduced for a broad range of noise frequencies by broadly suppressing both $F_\alpha(f)$ and $F_\Theta(f)$.

Notably, the predictions of Eqs.~\ref{eq:SF}-\ref{eq:angle_ff} match the simulated gate errors well at all noise frequencies, including those much lower than \mbox{$1/\tau = 6.7$ kHz}. At frequencies $f'$ lower than \mbox{1 kHz}, $\mathcal{E}_\Theta$ dominates $\mathcal{E}_\alpha$, as well as converges to a nonzero value as $f'\rightarrow 0$, which agrees with the simulated gate errors. Therefore, it is crucial to minimize $F_\Theta(f')$ in order to achieve robustness to noise that primarily occurs in the low-frequency regime, such as the $1/f$ noise. At frequencies $f'$ higher than \mbox{3 kHz}, $\mathcal{E}_\alpha$ dominates, so minimizing $F_\alpha(f')$ becomes crucial. 

Next, we show that the FF optimization reduces the gate error in the presence of mode-frequency noise of various spectrum. Instead of monotone fluctuations, we inject fluctuations $\delta(t)$ with respect to $S_\delta(f)$ in Eqs.~\ref{eq:Sdelta}-\ref{eq:Sdelta2} of various values of $f_c$. The robust-FM pulse is fixed to the one shown in Fig.~\ref{fig:pulses}, while the FF optimization is performed for each $f_c$ using the cost function in Eq.~\ref{eq:costfn}.

For the state-vector simulations, $\delta(t)$ \mbox{$(0 \leq t \leq \tau)$} is realized by assigning random phase to $S_\delta(f)^{1/2}$ independently at each frequency component and then performing an inverse Fourier transform. Each simulated gate error is averaged over 1000 realizations of noise.

Figure~\ref{fig:timevarying}(b) shows that the FF-optimized pulses have lower gate error than the robust-FM pulse for all noise PSDs considered. In most cases, the gate error is lower by more than an order of magnitude.

For noise PSDs of $f_c$ lower than 2 kHz, discrepancies between the simulated and predicted gate errors occur. This is because, as explained in Sec.~\ref{sec:subsec2D}, when noise is stronger at frequencies much lower than $1/\tau$, the first-order approximation of the FF formalism is less accurate \cite{Milne20, Kabytayev14}. 

In the regime where low-frequency noise is strong, gate errors are more accurately described when the noise is modeled as static parameter offsets~\cite{Kabytayev14}. For $f_c$ lower than \mbox{2 kHz}, Fig.~\ref{fig:timevarying}(b) also shows the average gate errors when the static offsets $\delta$, drawn from a normal distribution of zero mean and standard deviation \mbox{$2\pi \times (500^2 + 100^2)^{1/2}$ Hz}, are added to the mode frequencies. Each gate error is averaged over 1000 samples of $\delta$. When $f_c$ is lower than approximately \mbox{1 kHz}, these predictions better match the simulated gate errors than Eqs.~\ref{eq:SF}-\ref{eq:angle_ff} that use the FFs. However, for $f_c$ higher than \mbox{1 kHz}, predictions using the FFs show good match with the simulated gate errors. 

When a pulse is found by FF optimization combined with methods that achieve robustness to  static mode-frequency offsets beyond first order \cite{Kang21, Blumel21P}, the discrepancy between the simulated gate error and the prediction using the FFs can be removed. Such pulse can achieve even lower gate error in the presence of low-frequency noise. See Appendix~\ref{app:Batch} for an example. 

%%%%%%%%%%%%%%%FIGURE 3%%%%%%%%%%%%%%
\begin{figure*}[ht!]
\includegraphics[width=18.5cm]{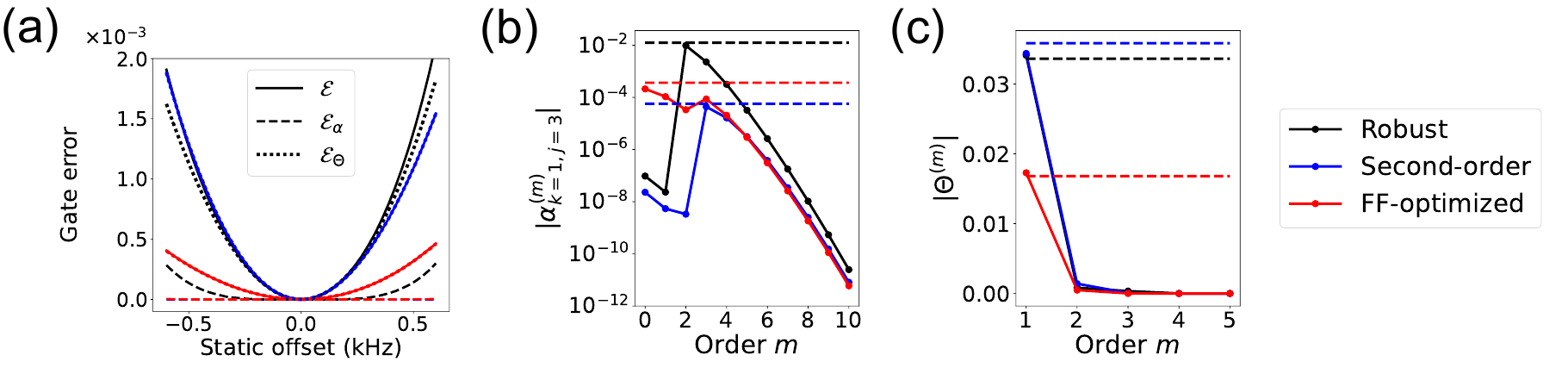}
\caption{(a) Simulated gate errors of the robust-FM (black), second-order robust-FM (blue), and FF-optimized (red) 150-$\mu$s pulses over various static offsets of $\omega_k$, applied uniformly for all $k$. The carrier Rabi frequencies of the three pulses are $\Omega/2\pi =$ 85.8, 96.6, and 109.8 kHz, respectively. The displacement errors $\mathcal{E}_\alpha = \mathcal{E} - \mathcal{E}_\Theta$ of the second-order robust-FM and FF-optimized pulses are too small to be visible. (b, c) Contributions of various orders to the (b) displacement [Eq. \ref{eq:alpham}] and (c) angle [Eq. \ref{eq:Thetam}] of the three pulses. Dashed lines show $|\alpha_{k=1,j=3}|$ and $|\Theta - \pi/4|$ when $\omega_k \rightarrow \omega_k + \delta_0$, where $\delta_0 = -2\pi \times 500$ Hz. In (c), the gap between the black and blue curves is too small to be visible.}
\label{fig:static}
\end{figure*}
%%%%%%%%%%%%%%%%%%%%%%%%%%%%%%%%%%%%

\subsection{Comparison under static offsets} \label{sec:subsec3C}

We show that suppressing the FFs at low frequencies achieves improved robustness to static mode-frequency offsets, as suggested in Sec.~\ref{sec:subsec2D}. Here, we compare the FF-optimization method with both the first-order and second-order robust-FM methods. For the first-order robust-FM and the FF-optimization methods, the pulses in Fig.~\ref{fig:pulses} are used. The second-order robust-FM pulse is found using the same parameters.

Figure~\ref{fig:static}(a) shows that in the presence of static offsets, the FF-optimized pulse achieves significantly lower gate error than the other two pulses. As in the case of low-frequency noise, the angle error dominates the displacement error~\footnote{The observation that the angle error dominates the displacement error does not apply generally to all MS gates. For example, Ref.~\cite{Sutherland22} shows that for a pulse of constant $\Omega$ and $\mu$ on a two-ion chain (and a single mode considered), the displacement error dominates the angle error, and even more so at higher temperatures.}, and the majority of the FF-optimized pulse's advantage comes from reducing the angle error. The displacement error of the FF-optimized pulse is also smaller than that of the first-order robust-FM pulse. 

As shown in Eq.~\ref{eq:staticFF}, $F_\alpha(f)$ contains the derivatives of the displacement over static offset of all orders. Therefore, the FF optimization is expected to suppress the higher-order derivatives, beyond the first-order derivative that is explicitly added to the cost function of Eq.~\ref{eq:costfn}.

To verify this, we calculate the terms of various orders in the displacement error of Eq.~\ref{eq:staticdisperr} in the presence of static offset $\delta_0 = -2\pi \times 500$ Hz, given by
\begin{equation} \label{eq:alpham}
    |\alpha^{(m)}_{kj}| = \left|\frac{\delta_0^m}{m!} \frac{\partial^m \alpha_{kj}}{\partial \omega_k^m} \right|.
\end{equation}
We show the case for $k=1$ and $j=3$, where $k=1$ corresponds to the mode of lowest frequency, as this displacement is the largest.

Figure~\ref{fig:static}(b) shows the result. As expected, for the first-order (second-order) robust-FM pulse, the second-order (third-order) term is the leading contribution to the displacement. Meanwhile, for the FF-optimized pulse, the 0th-order term, which is the leading contribution, is suppressed to a reasonably small value, and the terms up to fourth-order are ``evenly'' suppressed to within an order of magnitude smaller than the 0th-order term. This agrees with the intuition that the FF optimization suppresses the higher-order derivatives ($m>1$) when all terms of order lower than $m$ are suppressed to sufficiently small values, such that the summands in Eq.~\ref{eq:staticFF} are comparable. In particular, the second- and third-order terms of the FF-optimized pulse are more than an order of magnitude smaller than those of the first-order robust-FM pulse. This explains why the FF-optimized pulse has lower displacement errors than the first-order robust-FM pulse in Fig.~\ref{fig:static}(a). 

If the displacement error due to static offsets is the only source of error, then the second-order robust-FM pulse outperforms the FF-optimized pulse, as shown by the dashed lines of Fig.~\ref{fig:static}(b). Indeed, a shortcoming of the FF optimization is that completely removing $|\partial^m \alpha_{kj} / \partial \omega_k^m|$ is not guaranteed for any $m$. However, in practice, suppressing $|\partial^m \alpha_{kj} / \partial \omega_k^m|$ to small but nonzero values may be sufficient, as potentially larger contributions to gate error, such as the angle error and the displacement error due to time-varying mode-frequency fluctuations, can be additionally suppressed.

Now we consider the derivatives of the rotation angle over static mode-frequency offset. Figure~\ref{fig:static}(c) compares the $m$th-order contribution of static offsets \mbox{$\omega_k \rightarrow \omega_k + r_k \delta$} to the angle when $\delta = \delta_0$, i.e.,
\begin{equation}\label{eq:Thetam}
    |\Theta^{(m)}| = \left|\frac{\delta_0^m}{m!}  \frac{\partial^m \Theta}{\partial \delta^m} \right|,
\end{equation}
for various values of $m$. We again use $\delta_0 = -2\pi \times 500$ Hz. As robustness of the angle is not imposed for the first- and second-order robust-FM pulses, the FF-optimized pulse has about 2 times smaller $|\Theta^{(1)}|$ than these pulses, which leads the gate error to be about 4 times smaller. The higher-order contributions are negligible. 

A naturally arising question is whether $|\partial^m \Theta / \partial \delta^m|$ ($m=1,2,..$) can be completely removed, using reasonable pulse resources $\tau$ and $\Omega$, similarly to how $|\partial^m \alpha_{kj} / \partial \omega_k^m|$ is completely removed using the robust-FM methods of various orders. We note that this may require pulse-modulation schemes other than FM, such as amplitude modulation or amplitude-and-frequency modulation, as discussed in recent works~\cite{Ruzic22, Jia22}.

%%%%%%%%%%%%%%%%%%%%%%%%%%%%%%%%%%%%%%%%%%%%%%%%%%%%%%%%%%%%%%%%%%%%%%%%%%%%%%%%

%%%%%%%%%%%%%%%FIGURE 4%%%%%%%%%%%%%%
\begin{figure*}[ht!]
\includegraphics[width=18cm]{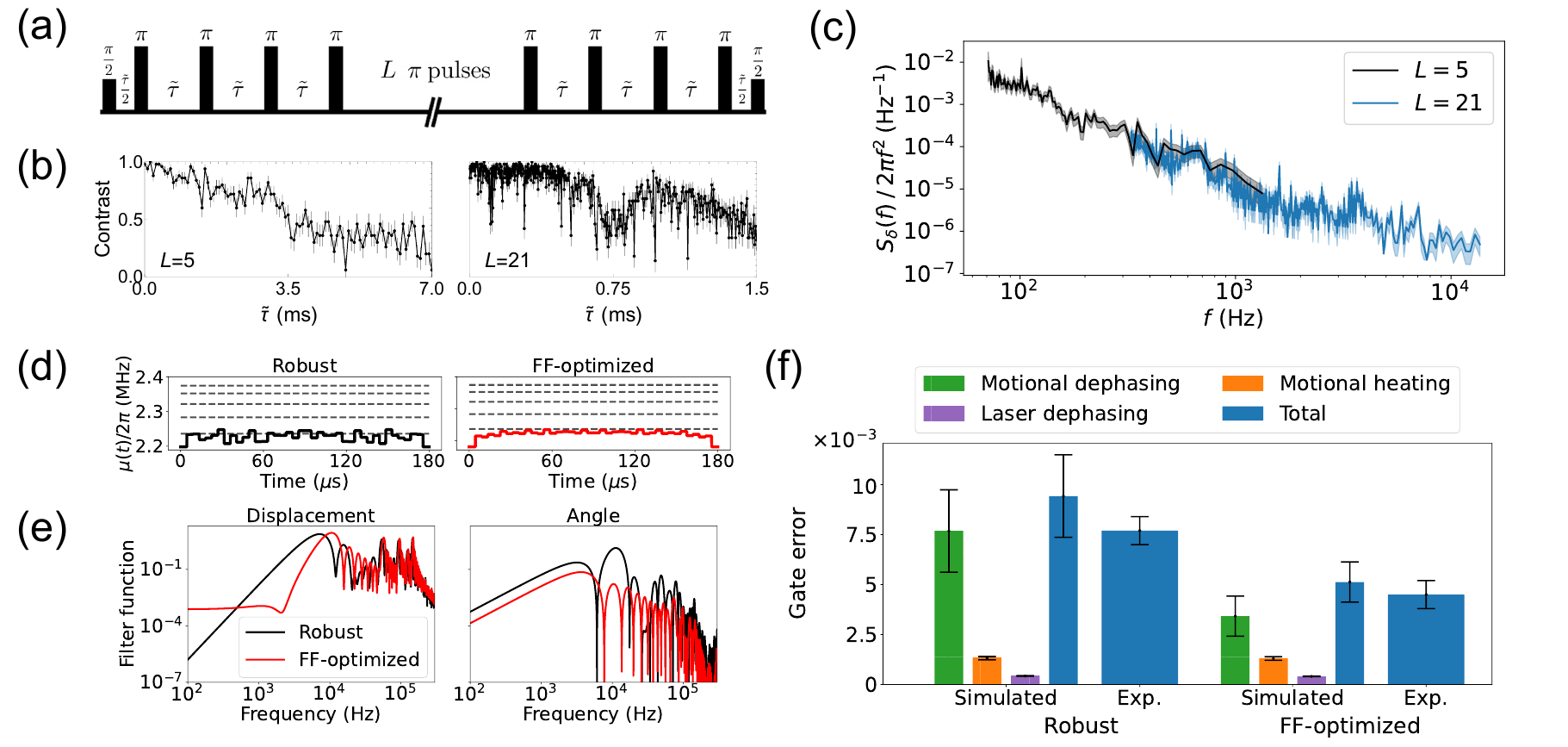}
\caption{Experimental data. (a) Diagram of the CPMG sequence. (b) Measured Ramsey contrast over various intervals between the $\pi$ pulses of the CPMG sequence. (c) Noise spectrum obtained from (b). Note that $S_\delta(f) / f^2$ is plotted in order to match Eq.~\ref{eq:SF}. The boundaries of the shaded region represent the standard deviation of the measured $S_\delta(f) / f^2$. (d) Robust-FM (left) and FF-optimized (right) pulses used in the experiments, which require carrier Rabi frequency \mbox{$\Omega/2\pi = $ 63.9} and 70.5 kHz, respectively. 180-$\mu$s pulses are applied on the second and third ions of a five-ion chain. The sideband frequencies are shown as dashed lines. (e) FFs $F_\alpha(f)$ (left) and $F_\Theta(f)$ (right) of the pulses. The FF optimization is performed using the noise spectrum measured in (c) with $L=21$. (f) Simulated budgets and experimentally measured values of the gate errors of the pulses. The experimentally measured gate errors of the robust-FM and FF-optimized pulses are 0.77(7)\% and 0.45(7)\%, respectively, where the difference comes from the effects of motional dephasing.} 
\label{fig:exp}
\end{figure*}
%%%%%%%%%%%%%%%%%%%%%%%%%%%%%%%%%%%%

\section{Experiment} \label{sec:sec4}

We measure the noise spectrum of the motional-mode frequencies in a five-ion chain of $^{171}{\rm Yb}^+$, and apply the results to the FF optimization. Then, we verify that the FF-optimized pulse has a higher MS-gate fidelity than the robust-FM pulse for a fixed pulse length of \mbox{180 $\mu$s}. The experimental setup is described in detail in Ref.~\cite{Wang20}. We use a rf system-on-chip (ZCU111), driven by firmware from Sandia National Laboratories \cite{QSCOUT}, as the rf source for modulating the laser pulses. 

Following the method of Ref.~\cite{Wang17}, we can measure the PSD of motional dephasing. We apply a CPMG sequence \cite{CP, MG} using the blue-sideband transition of the target motional mode and measure the Ramsey contrast at the end of the sequence. Then, the noise PSD $S_\delta(f)$ is obtained from the relations \cite{Wang17}
\begin{gather}
    \chi(\tilde{\tau}) = 4 \int^\infty_0 S_\delta(f) |\tilde{y}(f, \tilde{\tau})|^2 df,\\
    \tilde{y}(f, \tilde{\tau}) = \frac{1}{2\pi f} \sum_{j=0}^L (-1)^j 
    (e^{2\pi i f \tilde{\tau}_j} - e^{2\pi i f \tilde{\tau}_{j+1}}),
\end{gather}
where $e^{-\chi(\tilde{\tau})}$ is the Ramsey contrast, $L$ is the number of $\pi$ pulses in the CPMG sequence, $\tilde{\tau}$ is the interval time between $\pi$ pulses,  \mbox{$\tilde{\tau}_0 = 0$}, \mbox{$\tilde{\tau}_{L+1} = L \tilde{\tau}$}, and $\tilde{\tau}_i$ (\mbox{$i = 1,...,L$}) is the time stamp of the $i$th $\pi$ pulse, as shown in Fig.~\ref{fig:exp}(a). 
Note that $|\tilde{y}(f, \tilde{\tau})|^2$ can be interpreted as the filter function of the CPMG sequence.

The PSD measured with this method consists of dephasing in both spin control and the motional mode. Note that the coherence time of spin control in the system is close to $500$ ms, which is much longer than the $8$-ms motional-coherence time, so the measured PSD is dominated by motional dephasing. 

The measured Ramsey contrast and the noise PSD are shown in Fig.~\ref{fig:exp}(b) and (c), respectively. 
The noise spectrum is measured at frequencies below \mbox{14 kHz}, which is limited by the maximum available sideband-Rabi frequency of our system. We note that the methods in Refs.~\cite{Milne21, Keller21} may allow a wider bandwidth. 

Next, using the measured $S_\delta(f)$, we perform the FF optimization to find a MS-gate pulse, and compare with the robust-FM method. Figure~\ref{fig:exp}(e) shows that both $F_\alpha(f)$ and $F_\Theta(f)$ of the FF-optimized pulse are lower than those of the robust-FM pulse at most frequencies below \mbox{9 kHz}.

Finally, we experimentally measure the gate errors of the robust-FM and FF-optimized pulses. After initializing the qubits to $\ket{0}$, we apply sequences of various odd numbers of MS gates, which ideally generates the maximally entangled state \mbox{$(\ket{00} \pm i \ket{11}) / \sqrt{2}$}. In order to remove crosstalk errors, in a sequence of $2n+1$ gates ($n \geq 1$), we apply decoupling pulses after the first and second blocks of $n$ gates~\cite{Fang22}. The gate error $\mathcal{E}$ is extracted from a linear fit of the state errors, each given by \mbox{$\epsilon = \frac{1}{2}(p_{01} + p_{10} + 1 - c)$}, where $p_{01} + p_{10}$ is the populations of the $\ket{01}$ and $\ket{10}$ states combined and $c$ is the parity contrast \cite{Leibfried03}; see Appendices~\ref{app:GateFidMeas} and \ref{app:Crosstalk} for details. 

The measured MS-gate fidelity is 99.23(7)\% for the robust-FM method and 99.55(7)\% for the FF-optimized method, for a fixed pulse length of \mbox{180 $\mu$s}. The dominant sources of errors are motional dephasing, motional heating, and laser dephasing. Motional dephasing is simulated using the measured $S_\delta(f)$, and motional heating and laser dephasing are simulated using a master equation~\cite{Gardiner04}; see Appendix~\ref{app:ErrorBudget} for details. The error budget in Fig.~\ref{fig:exp}(f) shows that the error due to motional dephasing is reduced by more than half when the FF-optimized pulse is used. This demonstrates that noise in the mode frequencies can be characterized and then filtered out by designing the FFs, leading to an improved gate fidelity. 

We note in passing that our method of measuring the gate fidelity is specific to the initial qubit state $\ket{00}$. The gate fidelity may vary significantly for different initial states; see, for example, Ref.~\cite{Sutherland22}. Nonetheless, we expect that the FF optimization improves the gate fidelity for all initial states, as the FFs are designed to suppress the gate errors in Eqs.~\ref{eq:disp_err} and \ref{eq:angle_err}, which are derived from a distance measure between unitary operators~\cite{Bentley20} that is independent of the initial state. 

Our experimental setup allows limited laser intensity, which sets an upper bound on $\Omega$ at approximately \mbox{$2\pi \times 70$ kHz}. To meet this constraint, we use a cost function slightly modified from Eq.~\ref{eq:costfn} for the FF optimization. When $\Omega$ is upper bounded, for a shorter pulse length, designing a pulse that minimizes \mbox{$\sum_k \left( |\bar{\alpha}_{kj_1}|^2 + |\bar{\alpha}_{kj_2}|^2 \right)$} while satisfying $\Theta = \pi/4$ is more restrictive, which leaves smaller room for appropriately designing the FFs. For instance, when the pulse length is \mbox{150 $\mu$s}, the FF-optimized pulse does not outperform the robust-FM pulse with the current constraint on $\Omega$. Appendix \ref{app:ExpLimits} shows that if larger $\Omega$ is allowed, we expect a larger advantage of using the FF-optimized pulse even when an optimal pulse length is considered, 

Finally, we note that while $S_\delta(f)/f^2$ in Fig.~\ref{fig:exp}(c) roughly follows a $1/f^3$ trend, distinct high-frequency peaks have been observed in noise of other labs, such as the mode-frequency noise of Ref.~\cite{Milne20} and the laser-phase noise of Ref.~\cite{Nakav22}. In the presence of such fast noise, the FF optimization is expected to more significantly improve the gate fidelity, as shown in Fig.~\ref{fig:timevarying}.

%%%%%%%%%%%%%%%%%%%%%%%%%%%%%%%%%%%%%%%%%%%%%%%%%%%%%%%%%%

\section{Outlook and Conclusion} \label{sec:sec5}

While we demonstrate robustness to parameter fluctuations with respect to a first-order PSD, actual noise can be more complex. Spectroscopy tools for noise of higher-order spectrum \cite{Norris16, Sung19}, quantum noise \cite{Norris16}, and nonstationary noise \cite{Chalermpusitarak21} have been developed; however, the control and noise in these protocols have been limited to qubits. Whether such advanced noise spectroscopy can be used for oscillator-mediated entangling operations such as the MS gate is an interesting theoretical question.

We show that designing the FFs can improve the MS-gate fidelity in the presence of both time-varying fluctuations and static offsets of an experimental parameter. In general, we expect that the workflow of characterizing and filtering noise using the FF formalism will be useful for high-fidelity operations in trapped-ion systems, as well as various other quantum computing platforms. 

%%%%%%%%%%%%%%%%%%%%%%%%%%%%%%%%%%%%%%%%%%%%%%%%%%%%%%%%%%%%%%%%%%%%

\begin{acknowledgments}
This work is supported by the Office of the Director of National Intelligence, Intelligence Advanced Research Projects Activity through ARO Contract W911NF-16-1-0082, the National Science Foundation Expeditions in Computing Award 1730104, the National Science Foundation STAQ Project Phy-181891, and the U.S. Department of Energy, Office of Advanced Scientific Computing Research QSCOUT program.
\end{acknowledgments}

\appendix

%%%%%%%%%%%%%%%%%%%%%%%%%%%%%%%%%%%%%%%%%%%%%%%%%%%%%%%%%%%%%%%%%%%%%%%%%%%%%%

\section{ROBUSTNESS TO LASER-PHASE NOISE}
\label{app:Phase}

The FFs introduced in Eqs.~\ref{eq:disp_ff} and \ref{eq:angle_ff} can be designed to achieve robustness to time-varying fluctuations in the motional-mode frequencies. The same FFs can be used to achieve robustness to time-varying fluctuations in the laser phase.

For MS gates that use Raman beam pairs, motion phase and spin phase are defined from the phase differences of the beams. Depending on the orientation of the lasers, either motion phase or spin phase is chosen to be insensitive to the beam-path fluctuations \cite{Lee05}. Here, we consider the \textit{phase-insensitive} scheme, where the spin phase is insensitive and the motion phase's fluctuation is denoted as $\phi(t)$. We note that this does not apply to our experimental setup, which uses the phase-sensitive scheme. 

The lasers' motion-phase fluctuation $\phi(t)$ directly adds to the phases of the motional modes, i.e., \mbox{$\theta_k(t) \rightarrow \theta_k(t) + \phi(t)$} $\forall \: k$. This leads to the MS-gate error, with the components $\mathcal{E}_\nu$ ($\nu = \alpha, \Theta$) given by
\begin{equation}\label{eq:SFphi}
    \mathcal{E}_\nu = \int_{-\infty}^\infty df S_\phi(f) F_\nu(f),
\end{equation}
where $S_\phi(f)$ is the PSD of $\phi(t)$ defined analogously to Eq.~\ref{eq:Sdef}, and $F_\nu(f)$'s are found in Eqs.~\ref{eq:disp_ff} and \ref{eq:angle_ff} with $r_k = 1$ $\forall \: k$.

%%%%%%%%%%%%PHASE FIGURE%%%%%%%%%%%
\begin{figure}[ht!]
\includegraphics[width=8cm]{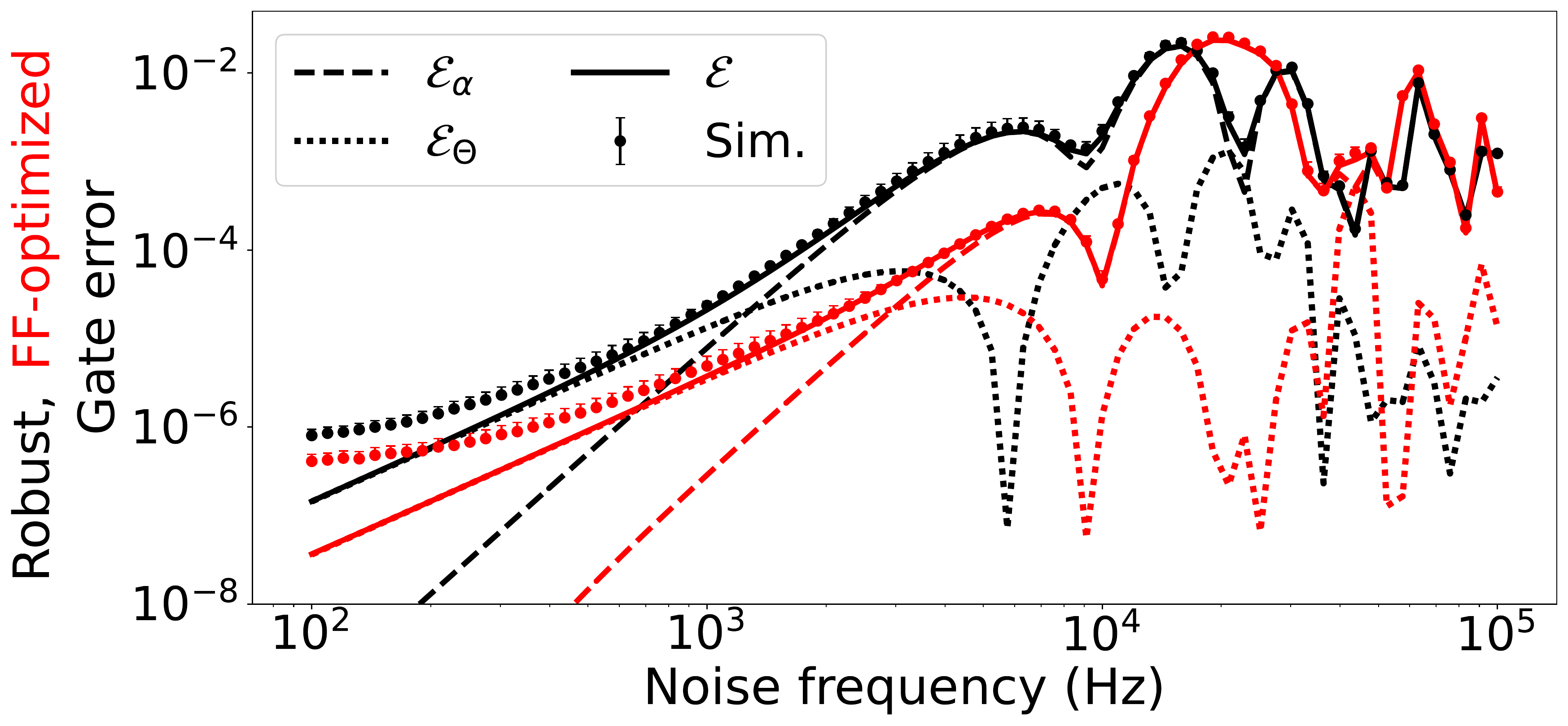}
\caption{Gate errors of the robust-FM (black) and FF-optimized (red) pulses with monotone noise of frequency $f'$ in the lasers' motion phase, for various values of $f'$. The amplitude of fluctuation is fixed to \mbox{$2\sqrt{2}\pi \times 0.01$ rad}. The gate errors are predicted (lines) by Eqs.~\ref{eq:SFphi},~\ref{eq:disp_ff}-\ref{eq:angle_ff}, and simulated (dots) by state-vector evolution. Each error bar represents the upper standard deviation of the simulated gate errors over 1000 initial phases of the fluctuation.}
\label{fig:phase}
\end{figure}
%%%%%%%%%%%%%%%%%%%%%%%%%%%%%%%%%%%%

%%%%%%%%%LASER INTENSITY FIGURE%%%%%
\begin{figure*}[ht!]
\includegraphics[height=5cm]{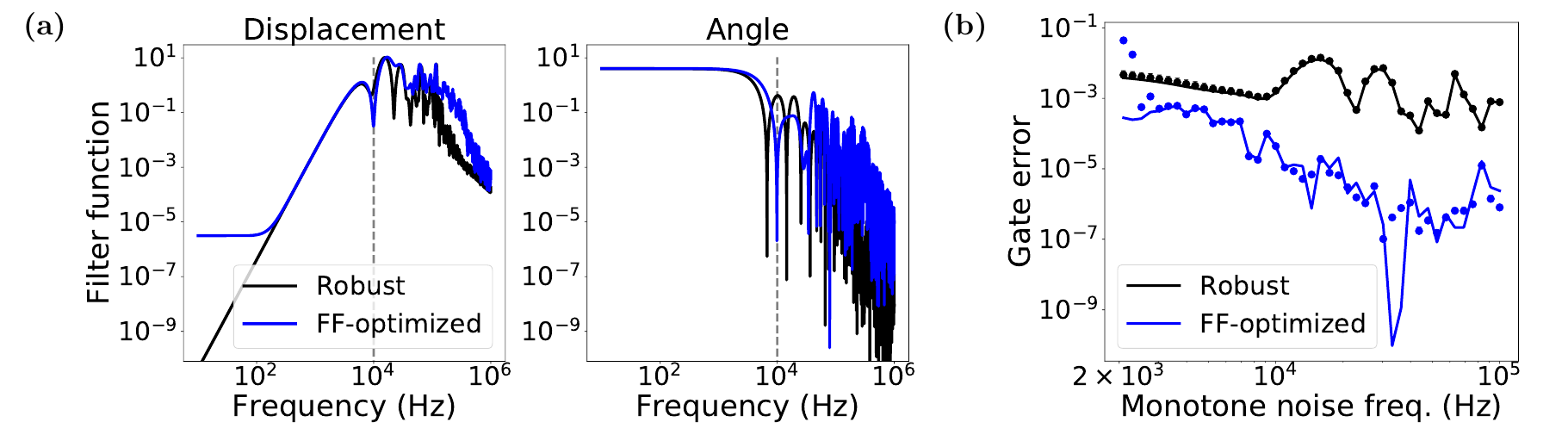}
\caption{(a) Filter functions $G_\alpha(f)$ (left) and $G_\Theta(f)$ (right) of the 150-$\mu$s pulses obtained by robust FM (black) and FF optimization (blue), which require carrier Rabi frequency $\Omega/2\pi =$ 85.8 and 118.5 kHz, respectively. Dashed lines show the monotone frequency \mbox{$f' = 10$ kHz} of the noise model used in the FF optimization. Note that the robust-FM pulse is not robust to laser-intensity offsets. (b) Gate errors with monotone noise of frequency $f'$ in the laser intensity, for various values of $f'$. The amplitude of carrier Rabi-frequency fluctuation is given by $A_{\Omega'} = \sqrt{2} \times 0.05 \Omega$. Each FF-optimized pulse, which requires carrier Rabi frequency $\Omega/2\pi$ between 66 and 285 kHz, is optimized with the corresponding monotone-noise PSD, and is compared with the robust-FM pulse shown in Fig.~\ref{fig:pulses}. The gate errors are predicted (lines) by Eqs.~\ref{eq:SFamp}-\ref{eq:angle_ff_amp}, and simulated (dots) by state-vector evolution. Each error bar, which represents the upper standard deviation of the simulated gate errors over 1000 initial phases the fluctuation, is too small to be visible.}
\label{fig:amp}
\end{figure*}
%%%%%%%%%%%%%%%%%%%%%%%%%%%%%%%%%%%%

Similarly to Fig.~\ref{fig:timevarying}(a), we inject monotone fluctuation of frequency $f'$ into the lasers' motion phase, and compare the gate errors predicted by Eqs.~\ref{eq:SFphi},~\ref{eq:disp_ff}-\ref{eq:angle_ff} and simulated using Qutip \cite{Qutip}, for various values of $f'$. We again use the robust-FM pulse and the FF-optimized pulse in Fig.~\ref{fig:pulses}.

Figure~\ref{fig:phase} shows the comparison. Similarly to Fig.~\ref{fig:timevarying}(a), the FF-optimized pulse has a lower gate error than the robust-FM pulse with any noise of frequency lower than \mbox{17 kHz}. This implies that robustness to fluctuations in the mode frequencies and those in the lasers' motion phase can be achieved simultaneously, as both parameters share the same filter functions. 

Unlike in Fig.~\ref{fig:timevarying}(a), the predicted gate errors do not perfectly match the simulated gate errors at low noise frequencies. In particular, Eq.~\ref{eq:SFphi} wrongly predicts that the gate error converges to zero as $f' \rightarrow 0$. It is expected that the FF formalism does not provide correct predictions for low-frequency noise in some parameters, as for noise of frequency much lower than $1/\tau$, the first-order approximation of the FFs is less accurate \cite{Milne20, Kabytayev14}. Nonetheless, $\mathcal{E}_\Theta$ provides a significantly closer match with the simulated gate errors than $\mathcal{E}_\alpha$ at low frequencies, which again highlights the relevance of designing $F_\Theta(f)$.

%%%%%%%%%%%%%%%%%%%%%%%%%%%%%%%%%%%%%%%%%%%%%%%%%%%%%%%%%%%%%%%%%%%%%%%%%%%%%%

\section{ROBUSTNESS TO LASER-INTENSITY NOISE}
\label{app:Laser}

The main text and Appendix~\ref{app:Phase} show that the FFs in Eqs.~\ref{eq:disp_ff} and \ref{eq:angle_ff} can be designed to achieve robustness to time-varying mode-frequency fluctuations $\delta_k(t)$ and laser-phase fluctuations $\phi(t)$. Here, we show that robustness to time-varying laser-intensity fluctuations, manifested as fluctuations in the carrier Rabi frequency, can also be achieved, but with a different set of FFs. 

We define $\Omega'(t)$ as the unintended fluctuations in $\Omega$, i.e. $\Omega \rightarrow \Omega + \Omega'(t)$. Then, similarly to Eqs.~\ref{eq:SF}-\ref{eq:angle_ff}, the MS-gate error terms $\mathcal{E}_\nu$ ($\nu = \alpha, \Theta$) due to $\Omega'(t)$ are given by

\begin{equation}
    \mathcal{E}_\nu = \int_{-\infty}^\infty df \frac{S_{\Omega'}(f)}{\Omega^2} G_{\nu} (f) \label{eq:SFamp},
\end{equation}
where
\begin{equation}
    G_\alpha(f) = \frac{\Omega^2}{4} \sum_k (\eta_{kj_1}^2 + \eta_{kj_2}^2)
                  \Big| \int_0^\tau dt \: e^{i(2\pi f t - \theta_k(t))} \Big|^2, \label{eq:disp_ff_amp}
\end{equation}
\begin{align}
    G_\Theta(f) &= \frac{\Omega^4}{4} \Big| \int^\tau_0 dt_1 \int^{t_1}_0 dt_2 \: 
    (e^{2\pi i f t_1} + e^{2 \pi i f t_2}) \nonumber\\
    &\quad\quad\quad\quad\:\times \sum_k \eta_{kj_1} \eta_{kj_2} \sin [\theta_k(t_1) - \theta_k(t_2)] \: \Big|^2. \label{eq:angle_ff_amp}
\end{align}
Here, $S_{\Omega'}(f)$ is the PSD of $\Omega'(t)$ defined analogously to Eq.~\ref{eq:Sdef}, and $G_\alpha(f)$ and $G_\Theta(f)$ are the FFs for the displacement and angle errors, respectively. These FFs can also be designed by performing an optimization with a cost function equivalent to Eq.~\ref{eq:costfn}.

The FFs for the displacement error $G_\alpha(f)$ turns out to be identical to $F_\alpha(f)$ with $r_k = 1$ (see Eq.~\ref{eq:disp_ff}). Therefore, suppressing the displacement error due to fluctuations in the mode frequency and the laser phase also suppresses that due to fluctuations in the laser intensity \cite{Milne20}. However, this does not hold for the angle error, as the FFs $G_\Theta(f)$ and $F_\Theta(f)$ (see Eq.~\ref{eq:angle_ff}) are different.

Figure~\ref{fig:amp}(a) shows the FFs $G_\alpha(f)$ and $G_\Theta(f)$ of the 150-$\mu$s pulses obtained by robust FM and FF optimization. The FF optimization is performed with a monotone noise PSD \mbox{$S_{\Omega'}(f) = (A_{\Omega'} / 2)^2 \times [\delta(f - f') + \delta(f + f')]$}, where \mbox{$A_{\Omega'} = \sqrt{2} \times 0.05 \Omega$} and \mbox{$f' = 10$ kHz} are the amplitude and frequency of the monotone fluctuation $\Omega'(t)$, respectively, and $\delta(\cdot)$ is the Dirac delta function. As expected, the FF-optimized pulse's FFs are both sharply suppressed at \mbox{$f = 10$ kHz}.

Note that unlike $F_\Theta(f)$, $G_\Theta(f)$ converges to a nonzero value as \mbox{$f \rightarrow 0$}. Indeed, when a static offset occurs in the carrier Rabi frequency from $\Omega$ to $\Omega + \Omega'$, the angle changes from $\Theta$ to $(1 + \Omega' / \Omega)^2 \times \Theta$, regardless of the pulse. Therefore, it is unlikely that $G_\Theta(f)$ is suppressed at zero or very low frequencies ($f \ll 1/\tau$) by pulse design. 

Figure~\ref{fig:amp}(b) shows the gate errors of the robust-FM pulse and the FF-optimized pulses with injected monotone laser-intensity noise. We note that the robust-FM pulse is robust to mode-frequency offsets but not to laser-intensity offsets. We vary the frequency $f'$ of the fluctuation of $\Omega'(t)$, where the amplitude of fluctuation is fixed to \mbox{$A_{\Omega'} = \sqrt{2} \times 0.05 \Omega$}. Similarly to Fig.~\ref{fig:timevarying}(b), the robust-FM pulse is fixed to the one shown in Fig.~\ref{fig:pulses}, while the FF optimization is performed for each monotone noise PSD. The pulse length is fixed to \mbox{150 $\mu$s}. 

For all noise frequencies higher than \mbox{2.5 kHz}, the FF-optimized pulses achieve significantly smaller gate error than the robust-FM pulse. However, at noise frequencies lower than 2.5 kHz, the first-order approximations of the FFs break down, and the FF optimization does not successfully reduce the gate error. 

Recently, Ref.~\cite{Shapira22} showed that robustness to static laser-intensity noise can be achieved by using spin-dependent squeezing. An interesting direction is to develop and design FFs for this gate scheme such that robustness to laser-intensity noise of any frequency is achieved. Also, one may consider \textit{circuit-level} error mitigation techniques \cite{Zhang22, Majumder22}, rather than gate-level pulse optimization, to handle static offsets and low-frequency fluctuations in the laser intensity.

%%%%%%%%%%%%%%%%%%%%%%%%%%%%%%%%%%%%%%%%%%%%%%%%%%%%%%%%%%%%%%%%%%%%%%%%%%%%%%

\section{GATE-FIDELITY MEASUREMENT}
\label{app:GateFidMeas}

\begin{figure}[ht!]
\includegraphics[width=8.6cm]{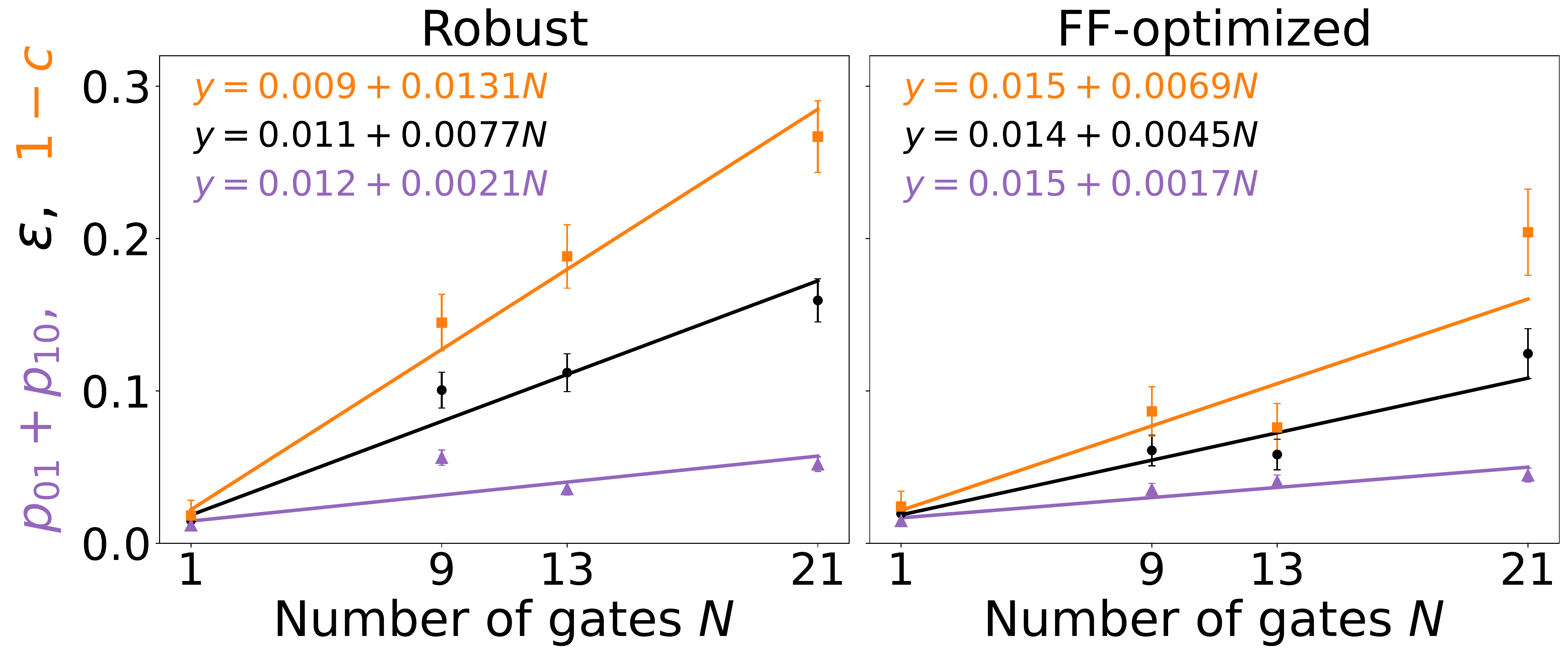}
\caption{Errors in the maximally entangled state generated by sequences of concatenated robust-FM (left) and FF-optimized (right) MS-gate pulses. The purple triangles, orange squares, and black circles are the population leakage to the $\ket{01}$ and $\ket{10}$ states, the loss of parity contrast, and the final-state errors, respectively. The gate error is extracted from the slope of the linear fit to the black circles.}
\label{fig:gatefidmeas}
\end{figure}

In this Appendix, we describe how the gate fidelities of the robust-FM and FF-optimized pulses are measured. We first initialize the qubits to $\ket{0}$, and then apply sequences of pulses, each consisting of 1, 9, 13, and 21 concatenated MS gates, which ideally generate the maximally entangled state \mbox{$(\ket{00} + i \ket{11}) / \sqrt{2}$}. Each sequence, except that with one gate, is interleaved with two pairs of single-qubit $Y$ gates, in order to mitigate optical crosstalk, as described in Appendix~\ref{app:Crosstalk}. The state error is given by \mbox{$\epsilon = \frac{1}{2}(p_{01} + p_{10} + 1 - c)$}, where $p_{01} + p_{10}$ is the measured populations of the $\ket{01}$ and $\ket{10}$ states combined and $c$ is the measured parity contrast \cite{Leibfried03}. Assuming that the coherent error is small, and using the fact that the stochastic error accumulates linearly and the state-preparation-and-measurement error remains constant with the number of concatenated gates, we find the gate fidelity $\mathcal{E}$ from the slope of a linear fit. According to the experimental data shown in Fig.~\ref{fig:gatefidmeas}, the measured MS-gate fidelity is 99.23(7)\% for the robust-FM method and 99.55(7)\% for the FF-optimized method, for a fixed pulse length of 180$\mu$s.

%%%%%%%%%%%GATE FID FIGURE%%%%%%%%%%
%%%%%%%%%%%%%%%%%%%%%%%%%%%%%%%%%%%%

Figure~\ref{fig:gatefidmeas} shows that the advantage of the FF-optimized pulse comes more from the smaller slope of the $1-c$ line than that of the $p_{01} + p_{10}$ line. This indicates that the lower gate error comes more from suppressing $\mathcal{E}_\Theta$ than $\mathcal{E}_\alpha$, because to leading order, \mbox{$p_{01} + p_{10} = \mathcal{E}_\alpha$} and \mbox{$1-c = \mathcal{E}_\alpha + 2\mathcal{E}_\Theta$} \cite{Milne20, Leibfried03}. This agrees with our observation that the measured noise spectrum $S_\delta(f)/f^2$ in Fig.~\ref{fig:exp}(c) is significantly larger at the low-frequency regime, and that $F_\Theta(f)$ dominates $F_\alpha(f)$ at low $f$. Therefore, suppressing $F_\Theta(f)$ is essential for achieving higher gate fidelities in the presence of low-frequency noise.

\section{ERROR-BUDGET SIMULATION}
\label{app:ErrorBudget}

In this section, we explain how the error budget for each pulse is evaluated in Fig.~\ref{fig:exp}(f). Table~\ref{tab:errorbudget} shows the data of Fig.~\ref{fig:exp}(f) in numbers. We consider the three most dominant sources of errors: motional dephasing, motional heating, and laser dephasing. Other sources, such as spontaneous emission and imperfection of the pulse solution, cause gate errors in the order of $10^{-4}$~\cite{Wang20}. Also, ac Stark shift, which is fourth order by design of our system with $^{171}{\rm Yb}^+$ ions, is carefully tracked such that its contribution to gate errors is negligible. 

\begin{table}[ht]
\renewcommand*{\arraystretch}{1.2}
\begin{tabular}{ c | c | c}
\hline
& Error of & Error of\\
Source of error & robust-FM pulse & FF-optimized pulse\\
& ($10^{-3}$) & ($10^{-3}$)\\
\hline
Motional dephasing & $7.7 \pm 2.1$ & $3.4 \pm 1.0$ \\
Motional heating & $1.3 \pm 0.1$ & $1.3 \pm 0.1$ \\
Laser dephasing & $0.43\pm 0.02$ & $0.41 \pm 0.02$\\
Total & $9.4 \pm 2.1$ & $5.1 \pm 1.0$\\
\hline
Experiment & $7.7 \pm 0.7$ & $4.5 \pm 0.7$ \\
\hline
\end{tabular}
\caption{Simulated budgets and experimentally measured values of the gate errors of the robust-FM and FF-optimized pulses used in the experiment. The data are identical to those shown in Fig.~\ref{fig:exp}(f).}\label{tab:errorbudget}
\end{table}

First, motional dephasing is simulated as noise in the mode frequencies according to $S_\delta(f)$ measured in Fig.~\ref{fig:exp}(c). Similarly to Sec.~\ref{sec:subsec3B}, fluctuation of mode frequencies $\delta(t)$ is realized in the time domain, by assigning random phase to $S_\delta(f)^{1/2}$ independently at each frequency component and then performing an inverse Fourier transform. Specifically, the value of $S_\delta(f)$ at each frequency used in simulations is drawn from a normal distribution of mean and standard deviation extracted from the measured $S_\delta(f)$. In the frequency region where measurements with $L=5$ and $L=21$ overlap, the average value of the two measurements is taken. 

For each $\delta(t)$, state-vector evolution is performed with respect to the Hamiltonian of the MS gate. Note that unlike in Sec.~\ref{sec:subsec3B}, $\delta(t)$ is generated for \mbox{$0 \leq t \leq 21\tau$}, such that the gate error is extracted from the slope of a linear fit of the state errors versus the numbers of concatenated gates, thus directly simulating the gate-fidelity-measurement experiment described in Appendix~\ref{app:GateFidMeas}. We use the simulated state errors averaged over 1000 realizations of noise, which follow a good linear trend. The simulated gate error due to motional dephasing and its uncertainty, shown in Fig.~\ref{fig:exp}(f) and Table~\ref{tab:errorbudget}, are the slope of the linear fit and its uncertainty, respectively.

Next, motional heating and laser dephasing are simulated using a master equation~\cite{Gardiner04}, following the method in the Supplemental Material of Ref.~\cite{Wang20}. The master equation is written in Lindblad form
\begin{equation*}
    \frac{d\hat{\rho}}{dt} = -i [\hat{H}, \hat{\rho}] + \sum_p \left(\hat{L}_p \hat{\rho} \hat{L}_p^\dagger - \frac{1}{2} \hat{L}_p^\dagger \hat{L}_p \hat{\rho} - \frac{1}{2} \hat{\rho} \hat{L}_p^\dagger \hat{L}_p\right),
\end{equation*}
where $\hat{\rho}$ is the density matrix, $\hat{H}$ is the Hamiltonian, and $\hat{L}_p$ is the $p$th Lindblad operator that describes its assigned decoherence process. Here, we consider a system consisting of two qubits $j_1$ and $j_2$ and one motional mode, truncated to the first ten Fock states. The evolution of each mode is simulated sequentially and then combined to obtain the final state, which relies on the fact that the residual entanglement between each mode and the qubits is small. Similarly as above, the state errors after concatenated MS-gate pulses are calculated, and then the gate error is extracted from the slope of a linear fit.  

Motional heating is described by the Lindblad operators \mbox{$\hat{L}_+ = \sqrt{\Gamma} \hat{a}^\dagger$} and \mbox{$\hat{L}_- = \sqrt{\Gamma} \hat{a}$}, where $\Gamma$ is the heating rate and $\hat{a}^\dagger$ is the creation operator of the mode. Laser dephasing is described by \mbox{$\hat{L}_l = \sqrt{1/T_l}(\hat{\sigma}^z_{j_1} + \hat{\sigma}^z_{j_2})$}, where $T_l$ is the laser coherence time and $\hat{\sigma}^z_{j}$ is the phase-flip operator of ion $j$. Based on experimental measurements, we use the heating rates \mbox{$\Gamma = 614(18)$ quanta/s} for the center-of-mass mode and \mbox{$\Gamma = 5$ quanta/s} for the other modes, and the laser coherence time \mbox{$T_l = 496(17)$ ms}~\cite{ZhangDissert}. The uncertainty of \mbox{$\Gamma$ ($T_l$)} leads to the uncertainty of the simulated gate error due to motional heating (laser dephasing) in Fig.~\ref{fig:exp}(f) and Table~\ref{tab:errorbudget}. 

Note that for the phase-insensitive laser orientation, the effects of laser dephasing can also be mitigated using the FFs as described in Appendix~\ref{app:Phase}. If the noise PSD of laser dephasing is known, laser dephasing can be simulated as fluctuation in an experimental parameter as well, instead of using a master equation. While motional heating cannot be mitigated using FFs, it can be significantly suppressed by, for instance, using a cryogenic system~\cite{Spivey21}. Overall, reducing the effects of motional dephasing, motional heating, and laser dephasing is a necessary step towards achieving high-fidelity gates with trapped ions.

\section{EXPERIMENTAL CONSTRAINTS OF PULSE OPTIMIZATION}
\label{app:ExpLimits}

For experimental implementation of the pulse optimization, there are several additional considerations.  First, the optimization should be performed within a few seconds, so that the runtime does not take a significant portion of the system's duty cycle. Second, certain modes are more susceptible to dissipative noise than other modes, so the drive frequency of the pulse needs to be far detuned from these modes. In the case of our experiment, the center-of-mass mode, which has the highest frequency, has a heating rate more than 100 times larger than that of the other modes. Third, the laser intensity is limited, which poses an upper bound on the carrier-Rabi frequency $\Omega$.

To satisfy the experimental constraints, we tweak the pulse optimization as the following. First, to reduce the run time of the FF optimization, instead of minimizing $F_\Theta(f)$ at various values of $f$, we minimize it only at the representative frequencies $f = \pm 1/2\tau$, where $\tau$ is the pulse length. This is because evaluating $F_\Theta(f)$ and its gradient is the most time-consuming routine at each iteration of optimization. We expect to improve the run time by, e.g., parallelization using graphics processing units. 

Second, to avoid exciting the center-of-mass mode, which is more than 100 times susceptible to heating than the other modes, we use an initial-guess pulse centered at the frequency \mbox{$\mu_0 = \min_k \omega_k - 2\pi \times 10$ kHz}. Given such an initial-guess pulse, both the robust-FM and FF-optimization methods are able to find pulses that are far detuned from the center-of-mass-mode frequency \mbox{$\max_k \omega_k$}, as shown in Fig.~\ref{fig:exp}(d). For longer ion chains with larger number of modes, the initial-guess pulse needs to be more carefully chosen, as different modes couple to different ions with varying strengths. 

Lastly, to find a pulse with a carrier Rabi frequency lower than the upper limit $\Omega_{\rm max}$, we add to the cost function a penalty term given by
\begin{equation}
    C_\Omega = \beta \exp \left\{ \gamma (1 - \Omega^2_{\rm max} / \Omega^2) \right\}, \nonumber
\end{equation}
where $\beta$ is chosen as $10^{-5}$ and $\gamma$ is typically chosen between 20 and 50. As $\gamma$ is large, $C_\Omega$ is very small when \mbox{$\Omega < \Omega_{\rm max}$} but becomes large when \mbox{$\Omega > \Omega_{\rm max}$}. This ensures $\Omega \lesssim \Omega_{\rm max}$ when the overall cost function is minimized. For the gate-fidelity-measurement experiment with the pulses in Fig.~\ref{fig:exp}(d), we use \mbox{$\Omega_{\rm max} = 2\pi \times 70$ kHz}.

To find the minimum achievable gate error for a given $\Omega_{\rm max}$, we perform simulations for various pulse lengths. In general, as the pulse length increases, the gate error tends to increase, as the effects of dissipative noise build up over time. However, when the pulse length is too short, a sufficiently good pulse solution that satisfies $\Omega \lesssim \Omega_{\rm max}$ cannot be found, so the gate error becomes larger. Therefore, there exists an optimal pulse length that achieves the lowest gate error for a given $\Omega_{\rm max}$.

%%%%%%%%%RABI LIMIT FIGURE%%%%%
\begin{figure}[ht!]
\includegraphics[width=8cm]{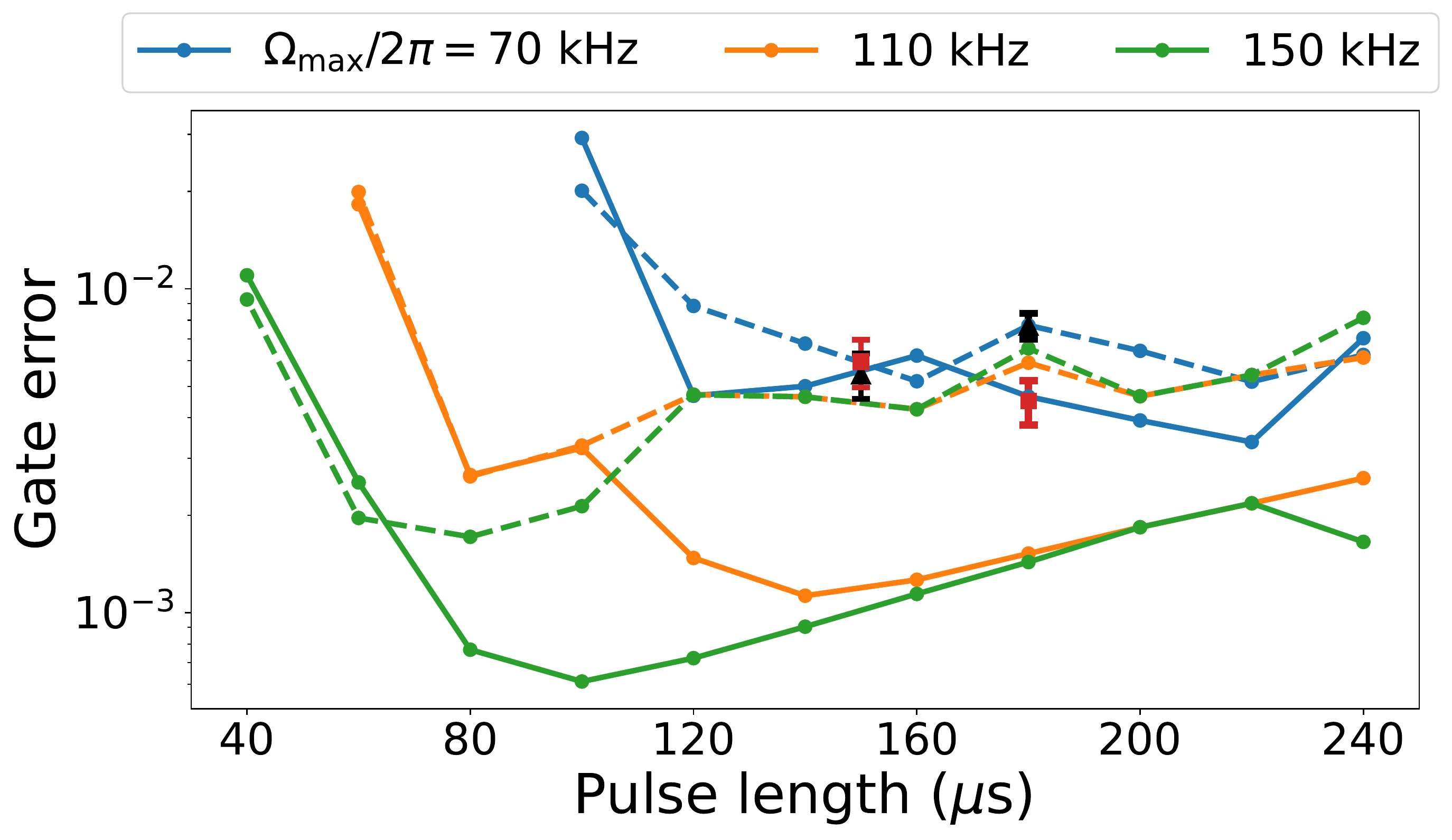}
\caption{Simulated gate errors of the robust-FM (dashed) and FF-optimized (solid) pulses, for various values of pulse length and upper bound on the carrier Rabi frequency. The experimentally measured gate errors of the robust-FM (FF-optimized) pulses of lengths 150 and \mbox{180 $\mu$s} found with \mbox{$\Omega_{\rm max} = 2\pi \times 70$ kHz} are marked as the black triangles (red squares).}
\label{fig:Rabilimit}
\end{figure}
%%%%%%%%%%%%%%%%%%%%%%%%%%%%%%%%%%%%

The simulated gate errors are shown in Fig.~\ref{fig:Rabilimit}. Similarly to Appendix~\ref{app:ErrorBudget}, each simulated gate error is obtained from a linear fit of the state errors versus the numbers of concatenated gates up to 21. Motional dephasing, motional heating, and laser dephasing are simulated altogether by solving a master equation with the mode frequencies fluctuating according to $S_\delta(f)$. We use the same values of noise parameters as in Appendix~\ref{app:ErrorBudget}. Each state error is averaged over 300 realizations of motional dephasing. 

We also show the experimentally measured gate errors of the robust-FM and FF-optimized pulses of lengths 150 and \mbox{180 $\mu$s} found with \mbox{$\Omega_{\rm max} = 2\pi \times 70$ kHz}, which match well with the simulated gate errors. The FF-optimized pulse outperforms the robust-FM pulse when the pulse length is \mbox{180 $\mu$s}, but not when the pulse length is \mbox{150 $\mu$s}. This is because when $\Omega$ has an upper limit, as the pulse length gets shorter, the condition of achieving high-fidelity MS gate \textit{without} noise becomes already more restrictive, which leaves smaller room for the FFs to be appropriately designed.

Figure~\ref{fig:Rabilimit} shows that when the optimal pulse length is considered, we expect a larger advantage of using the FF-optimized pulse when $\Omega_{\rm max}$ is larger. In particular, when \mbox{$\Omega_{\rm max} = 2\pi \times 150$ kHz}, the lowest simulated gate error of the robust-FM (FF-optimized) pulse is 0.17\% (0.061\%), when the pulse length is 80 (100) $\mu$s. Therefore, we expect the FF optimization to be even more useful in future experiments that allow larger laser intensity without introducing additional technical noise.

\section{CROSSTALK SUPPRESSION}
\label{app:Crosstalk}

Crosstalk errors need to be considered when implementing two-qubit gates in a chain of more than two ions, as the unwanted entanglement between the target and spectator ions created by crosstalk impacts the fidelity of the Bell state of the target ions. The crosstalk between target ion $i$ and spectator ion $j$ is quantified as the carrier-Rabi-frequency ratio $\epsilon_{ij} = \Omega_j/\Omega_i$ when resonantly driving a single-qubit gate on ion $i$. In our system we measure $\epsilon_{ij}$ to be $1 - 2\%$ for nearest neighbors due to imperfect optical addressing mainly caused by aberrations. The crosstalk level is within the range of state-of-the-art trapped-ion experiments, but this still needs to be mitigated in order to attain a $99.5 \%$-level two-qubit-gate fidelity. Due to the coherent nature of crosstalk, its effect can be actively canceled by applying single-qubit spin-echo pulses in the middle of the gate(s), reversing the crosstalk interaction during the second half of the MS evolution \cite{Parrado21, Fang22}. 

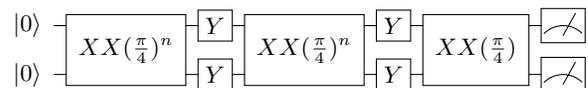
\begin{figure}[!htb]
\centerline{
\Qcircuit @C=0.5em @R=0.6em {
& \lstick{\ket{0}} & \multigate{1}{XX(\frac{\pi}{4})^{n}} & \gate{Y} & \multigate{1}{XX(\frac{\pi}{4})^{n}} & \gate{Y} & \multigate{1}{XX(\frac{\pi}{4})} & \meter \\
& \lstick{\ket{0}} & \ghost{XX(\frac{\pi}{4})^{n}} & \gate{Y} & \ghost{XX(\frac{\pi}{4})^{n}} & \gate{Y} & \ghost{XX(\frac{\pi}{4})} & \meter
}}
\caption{Circuit diagram of a crosstalk-suppression scheme for each sequence of $2n+1$ ($n \geq 1$) concatenated MS gates. The sequence is interleaved with two pairs of $Y$ gates on the target ions, such that the crosstalk interaction is reversed during the second sequence of $n$ gates.}
\label{fig:crosstalk}
\end{figure}

For each sequence of $2n+1$ $(n \geq 1)$ concatenated MS gates in the gate-fidelity measurement described in Appendix~\ref{app:GateFidMeas}, we use a crosstalk-suppression scheme that applies the echoing pulses on the \textit{target} ions, as illustrated in the circuit of Fig.~\ref{fig:crosstalk}~and detailed in Ref.~\cite{Fang22}. Note that a pair of $Y$ gates commutes with a MS gate, so the $Y$ gates would not affect the final state in ideal conditions. A single MS gate is applied after the second pair of $Y$ gates in order to generate the Bell state for the fidelity measurement.

\section{BATCH OPTIMIZATION OF FILTER FUNCTIONS}
\label{app:Batch}

When the frequency of noise is much lower than $1/\tau$, noise essentially becomes a static parameter offset within the duration of a single gate. In the FF optimization, which uses the cost function in Eq.~\ref{eq:costfn}, the first term minimizes the gate error due to static mode-frequency offsets up to first order. However, higher-order errors are not minimized, which causes the first-order approximation of the FF formalism to be less accurate. Indeed, the simulated gate errors are higher than the predictions using the FFs in Fig.~\ref{fig:timevarying}(b), when the low-frequency component of noise is relatively strong. This motivates combining the FF optimization with pulse-design methods that achieve robustness to static offsets of motional-mode frequencies beyond first order \cite{Blumel21P, Kang21}.

Here we combine FF optimization with the ``b(atch)-robust FM,'' introduced in Ref.~\cite{Kang21}. Instead of using an analytic robustness condition, the b-robust FM minimizes the average gate error over a range of systematic errors. When the batch size is 1, the cost function is given by
\begin{align}
    C(\vec{\delta}) &= 
    \sum_{j=j_1,j_2} \sum_k |\alpha_{kj}( \vec{\delta})|^2
    + \frac{1}{2}\big(\Theta(\vec{\delta}) - \frac{\pi}{4}\big)^2 \nonumber\\
    &+ \int_{-f_{\rm max}}^{f_{\rm max}} df \frac{S_\delta(f)}{f^2} [F_\alpha(f, \vec{\delta}) + F_\Theta(f, \vec{\delta})]. \nonumber
\end{align}
Here, $\vec{\delta}$ is the offset vector whose $k$-th element is $\delta_k$, and $\alpha_{kj}(\vec{\delta})$, $\Theta(\vec{\delta})$, and $F_\nu(f, \vec{\delta})$ are, respectively, the displacement, rotation angle, and filter function when $\omega_k$ is replaced by $\omega_k + \delta_k$. At each iteration of optimization, $\vec{\delta}$ is randomly updated, where each $\delta_k$ is drawn from a normal distribution of mean zero and standard deviation $2\pi \times 0.5$ kHz. The adaptive-moment-estimation \cite{ADAM} optimizer is used in order to stabilize the gradient while the cost function changes over iterations. 

%%%%%%%%%%%%%%%FIGURE 6%%%%%%%%%%%%%%
\begin{figure}[ht]
\includegraphics[height=4.5cm]{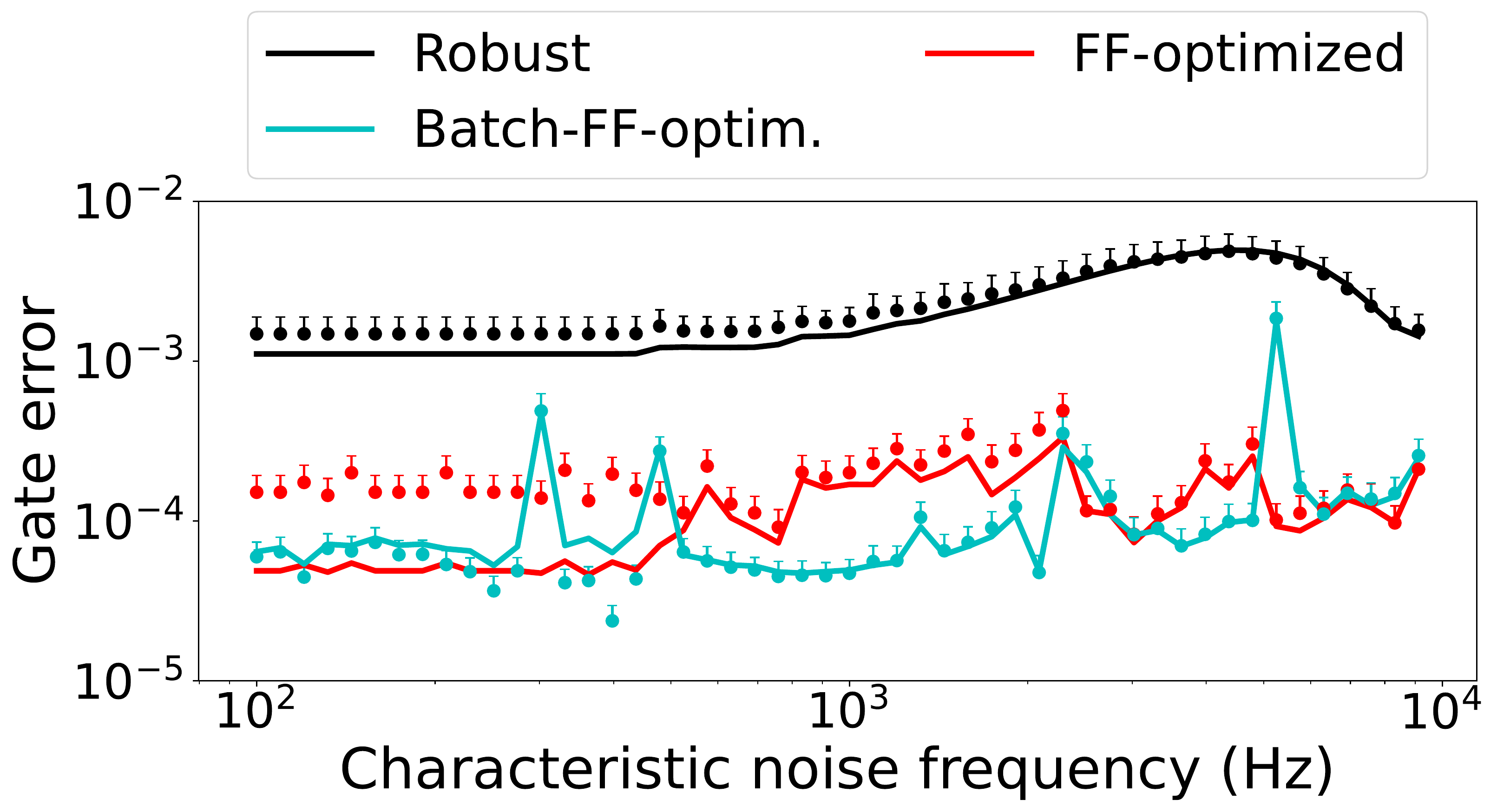}
\caption{Gate errors of the pulses obtained by batch-FF optimization under various noise of spectrum $S_\delta(f)$, each defined with the characteristic frequency $f_c$ by Eqs.~\ref{eq:Sdelta}-\ref{eq:Sdelta2}. Each batch-FF-optimized pulse, which requires carrier Rabi frequency $\Omega/2\pi$ between 90 and 150 kHz, is compared with the robust-FM and plain FF-optimized pulses used in Fig.~\ref{fig:timevarying}(b). The gate errors are predicted (lines) by Eqs.~\ref{eq:SF}-\ref{eq:angle_ff} and simulated (dots) by state-vector evolution. Each error bar represents the upper standard deviation of the simulated gate errors over 1000 realizations of noise.}
\label{fig:batch}
\end{figure}
%%%%%%%%%%%%%%%%%%%%%%%%%%%%%%%%%%%%

Similarly to Fig.~\ref{fig:timevarying}(b), this batch optimization is performed for various noise PSDs. To reduce the runtime, for the optimization we use the noise spectrum $S_\delta(f) = (2\pi \times 0.5\:{\rm kHz})^2 / 2 \times [\delta(f - f_c) + \delta(f + f_c)]$, where $\delta(\cdot)$ is the Dirac $\delta$ function. Figure~\ref{fig:batch} shows the simulated and predicted gate errors, compared with the pulses used in Fig.~\ref{fig:timevarying}(b) obtained by robust FM and FF optimization without batch. Except a few outliers, the batch-FF-optimized pulses have even lower gate error than the plain FF-optimized pulses. Furthermore, the match between the gate errors simulated by state-vector evolution and the gate errors predicted by Eqs.~\ref{eq:SF}-\ref{eq:angle_ff} is improved, especially with low $f_c$. This is because the batch optimization achieves robustness to static offsets of mode frequencies beyond first order. 

While the batch-FF optimization is promising especially with low-frequency noise, it takes significantly longer runtime than the plain FF optimization, as minimizing a randomly updated cost function requires a larger number of iterations. For the pulses in Fig.~\ref{fig:batch}, we perform 10000 iterations for each batch optimization using the adaptive-moment-estimation optimizer, while less than 300 iterations was sufficient for each plain optimization using the BFGS optimizer \cite{NumericalOptimization}. For experimental application of pulse optimization to a long ion chain, efficient and parallelized implementation of the algorithm should be accompanied. See Ref.~\cite{Kang21} for a discussion of typical runtimes.

\section{FLUCTUATIONS OF THE LAMB-DICKE PARAMETERS}

In this Appendix, we consider the fluctuations of the Lamb-Dicke parameters $\eta_{kj}$. The definition of $\eta_{kj}$ is given by
\begin{equation}
    \eta_{kj} = b_{kj} \Delta k \sqrt{\frac{\hbar}{2m_{\rm ion} \omega_k}}, \nonumber
\end{equation}
where $b_{kj}$ is the $j$-th element of the normalized eigenvector of the $k$-th normal mode with eigenvalue $\omega_k$, $\Delta k$ is the magnitude of the wave-vector difference of the two Raman beams along the motional direction of the modes, and $m_{\rm ion}$ is the mass of the ion.

Here we consider the fluctuations of the rf driving signal for the trap, which is the leading source of error in our experimental setup. Up to first order, the rf-voltage fluctuations are expected to uniformly ``scale up or down'' the mode frequencies, i.e., $\omega_k \rightarrow \omega_k (1 + \delta(t) / \omega_{\rm CM})$, and not affect the eigenvector elements $b_{kj}$. However, $\eta_{kj}$ still has an explicit proportionality to $\sqrt{1/\omega_k}$. While this does not affect the displacement error, as $\alpha_{kj}$ is proportional to $\eta_{kj}$ and is minimized to zero, this may affect the angle $\Theta$. Specifically, when $\omega_k \rightarrow \omega_k (1 + \delta(t) / \omega_{\rm CM})$, the $\eta_{kj_1} \eta_{kj_2}$ factor in $\Theta$ is scaled to $\eta_{kj_1} \eta_{kj_2} (1 + \delta(t) / \omega_{\rm CM})^{-1}$ (see Eq.~\ref{eq:Theta}). 

Therefore, rf-voltage fluctuations have two effects: (i) dephasing, or fluctuation of \mbox{$\theta_k(t) = \int_0^t [\mu(t') - \omega_k] dt'$}, and (ii) amplitude fluctuation due to \mbox{$\eta_{kj_1} \eta_{kj_2} \propto 1/\omega_k$}. We compare the magnitude of the two effects on $\Theta$ in the presence of static noise, i.e., \mbox{$\omega_k \rightarrow \omega_k (1 + \delta / \omega_{\rm CM})$}. Specifically, we compare $|\partial \Theta / \partial \delta|$ in Eq.~\ref{eq:anglederivative} when only one of the effects in (i) and (ii) exists. The effect of dephasing on $\Theta$ is reduced by the FF optimization, so we use the FF-optimized pulse in Fig.~1 as an example. When only dephasing exists, $|\partial \Theta / \partial \delta| = 5.15 \times 10^{-6}\:\text{Hz}^{-1}$. When only amplitude fluctuation exists, $|\partial \Theta / \partial \delta| = \Theta / \omega_{\rm CM} = 5.35 \times 10^{-8}\:\text{Hz}^{-1}$. Therefore, the effect of dephasing is 2 orders of magnitude larger than that of amplitude fluctuation. This justifies why we achieve only robustness to dephasing and not the fluctuation of $\eta_{kj}$. 

We note that for future pulse-optimization schemes that achieve more improved robustness of $\Theta$ to dephasing, the fluctuation of $\eta_{kj}$ may need to be considered. To achieve robustness to $\eta_{kj}$ fluctuation at any frequency, a possible method is to suppress $|\partial \Theta / \partial \delta|$ and $F_\Theta(f)$ derived for the case where $\eta_{kj}$ is an explicit function of $\omega_k$.

\section{DERIVATIONS OF THE ANGLE FF}
\label{app:Derivations}

In this Appendix, we present the derivations of Eq.~\ref{eq:SF} for $\nu = \Theta$ and Eq.~\ref{eq:angle_ff}, which define the angle FF $F_\Theta(f)$. The derivations for the displacement FF $F_\alpha(f)$ can be found in Refs.~\cite{Green15, Milne20}.

We consider a time-varying fluctuation $\varphi_k(t)$ in the phase $\theta_k(t)$ of motional mode $k$, such that \mbox{$\theta_k(t) \rightarrow \theta_k(t) + \varphi_k(t)$}. To first order in $\varphi_k(t)$, the angle $\Theta$ becomes
\begin{widetext}
\begin{align}
    \Theta &= - \Omega^2 \sum_k \frac{\eta_{kj_1} \eta_{kj_2}}{2} \int^\tau_0 dt_1 \int^{t_1}_0 dt_2 
    \sin [\theta_k(t_1) - \theta_k(t_2) + \varphi_k(t_1) - \varphi_k(t_2)] \nonumber \\
    &\approx - \Omega^2 \sum_k \frac{\eta_{kj_1} \eta_{kj_2}}{2} \int^\tau_0 dt_1 \int^{t_1}_0 dt_2 
    \Big( \sin [\theta_k(t_1) - \theta_k(t_2)] 
    + [\varphi_k(t_1) - \varphi_k(t_2)] \times \cos[\theta_k(t_1) - \theta_k(t_2)]  \Big). \nonumber
\end{align}
\end{widetext}
When $\varphi_k(t) = 0$ $\forall k$, $\Theta$ is equal to its ideal value $\pi/4$. For brevity, we assume that $\varphi_k(t) = r_k \varphi(t)$, i.e. dephasing of different modes differ only up to proportionality constants. The angle gate error $\mathcal{E}_\Theta$, given by Eq.~4, becomes
\begin{widetext}
\begin{align}
    \mathcal{E}_\Theta &= \left| 
    \frac{\Omega^2}{2}  
    \int^\tau_0 dt_1 \int^{t_1}_0 dt_2  [\varphi(t_1) - \varphi(t_2)]
    \sum_k r_k \eta_{kj_1} \eta_{kj_2} \cos[\theta_k(t_1) - \theta_k(t_2)]
    \right|^2 . \nonumber
\end{align}
\end{widetext}
Now we use $\mathbb{E}[\varphi(t)\varphi(t')] = \int^\infty_{-\infty} df S_\varphi(f) e^{2\pi i f (t - t')}$ from the definition of the PSD of the phase noise $S_\varphi(t)$, where $\mathbb{E}[\cdot]$ denotes the expectation value of the argument. Also note that $S_\varphi(t) = S_\delta(f)/f^2$, as $\varphi(t) = \int_0^t \delta(t')dt'$. Then, the expectation value of $\mathcal{E}_\Theta$ is given by
\begin{widetext}
\begin{align}
    \mathbb{E}[\mathcal{E}_\Theta] &= 
    \frac{\Omega^4}{4} \int^\tau_0 dt_1 \int^{t_1}_0 dt_2 \int^\tau_0 dt'_1 \int^{t'_1}_0 dt'_2 
    \int^\infty_{-\infty} df S_\varphi(f) 
    \Big(e^{2\pi i f (t_1 - t'_1)} - e^{2\pi i f (t_1 - t'_2)} - e^{2\pi i f (t_2 - t'_1)} + e^{2\pi i f (t_2 - t'_2)}\Big) \nonumber \\
    &\quad\quad\quad\quad\quad\quad\quad\quad\quad\quad\quad\quad\quad\quad\quad\quad
    \times \sum_{k,k'}  r_k r_{k'} \eta_{kj_1} \eta_{kj_2} \eta_{k'j_1} \eta_{k'j_2}
    \cos[\theta_k(t_1) - \theta_k(t_2)] 
    \cos[\theta_{k'}(t'_1) - \theta_{k'}(t'_2)] \nonumber\\
    &= \int^\infty_{-\infty} df \frac{S_\delta(f)}{f^2} F_\Theta(f), \nonumber
\end{align}
where
\begin{align}
    F_\Theta(f) &= 
    \frac{\Omega^4}{4} \int^\tau_0 dt_1 \int^{t_1}_0 dt_2 \int^\tau_0 dt'_1 \int^{t'_1}_0 dt'_2 
    \Big(e^{2\pi i f t_1} - e^{2\pi i f t_2}\Big) \Big(e^{-2\pi i f t'_1} - e^{-2\pi i f t'_2}\Big)
    \nonumber \\
    &\quad\quad\quad\quad\quad\quad\quad\quad\quad\quad\quad\quad\quad\quad
    \times \sum_{k,k'} r_k r_{k'} \eta_{kj_1} \eta_{kj_2} \eta_{k'j_1} \eta_{k'j_2}
    \cos[\theta_k(t_1) - \theta_k(t_2)] 
    \cos[\theta_{k'}(t'_1) - \theta_{k'}(t'_2)] \nonumber\\
    &= \Omega^4 \left|\int^\tau_0 dt_1 \int^{t_1}_0 dt_2 \: 
    (e^{2\pi i f t_1} - e^{2 \pi i f t_2})
    \sum_k \frac{r_k}{2}\: \eta_{kj_1} \eta_{kj_2} 
    \cos [\theta_k(t_1) - \theta_k(t_2)] \: \right|^2. \nonumber
\end{align}
\end{widetext}
This completes the derivation of Eq.~\ref{eq:SF} for $\nu = \Theta$ and Eq.~\ref{eq:angle_ff}, where we use $\mathcal{E}_\Theta$ instead of $\mathbb{E}[\mathcal{E}_\Theta]$ to denote the expected angle error in the presence of time-varying fluctuations. The angle FF for the laser intensity noise $G_\Theta(f)$, defined in Eq.~\ref{eq:SFamp} for $\nu = \Theta$ and Eq.~\ref{eq:angle_ff_amp}, can also be derived in a similar way. 

\bibliography{bib}% Produces the bibliography via BibTeX.
\end{document}